\documentclass[journal=ancac3,manuscript=article]{achemso}
\usepackage[version=3]{mhchem}

\usepackage{amsmath}
\usepackage{amssymb}
\usepackage{graphicx}
\usepackage{dcolumn}
\usepackage{bm}
\usepackage[mathlines]{lineno}
\usepackage{xcolor}
\usepackage{comment}

\title{Orbits, spirals, and trapped states: \\Dynamics of a phoretic Janus particle in a radial concentration gradient}

\author{Parvin Bayati}
\affiliation{Department of Chemistry, The Pennsylvania State University, University Park, Pennsylvania, 16802, USA}

\author{Stewart A. Mallory}
\email{sam7808@psu.edu}
\affiliation{Department of Chemistry, The Pennsylvania State University, University Park, Pennsylvania, 16802, USA}
\altaffiliation{Department of Chemical Engineering, The Pennsylvania State University, University Park, Pennsylvania, 16802, USA}


\begin{document}

\begin{tocentry}
\centering
\includegraphics[]{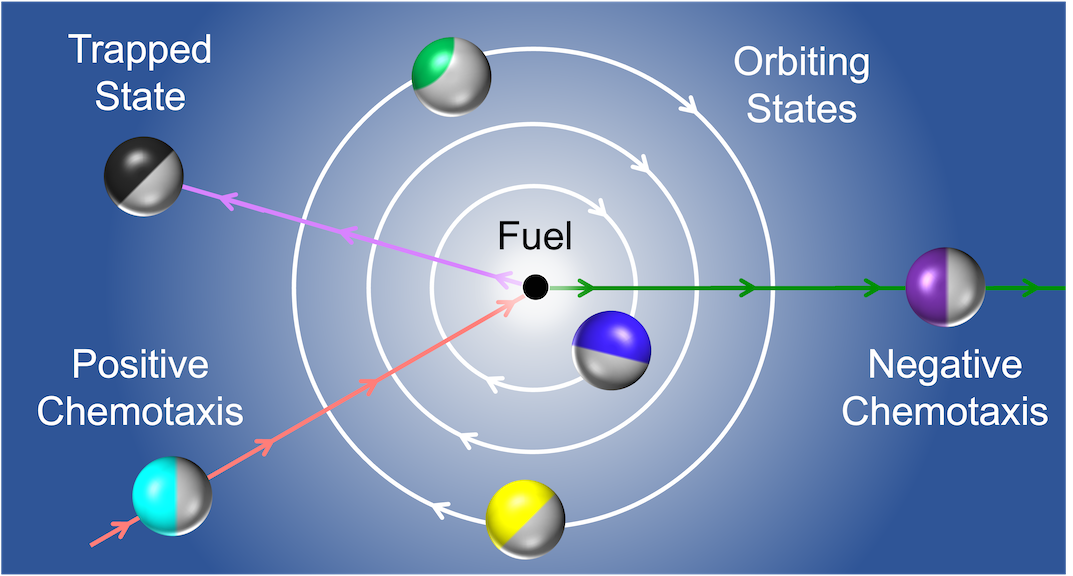}





\end{tocentry}

\begin{abstract}
A longstanding goal in colloidal active matter is to understand how gradients in fuel concentration influence the motion of phoretic Janus particles.
Here, we present a theoretical description of the motion of a spherical phoretic Janus particle in the presence of a radial gradient of the chemical solute driving self-propulsion.
Radial gradients are a geometry relevant to many scenarios in active matter systems and naturally arise due to the presence of a point source or sink of fuel.
We derive an analytical solution for the Janus particle's velocity and quantify the influence of the radial concentration gradient on the particle's trajectory.
Compared to a phoretic Janus particle in a linear gradient in fuel concentration, we uncover a much richer set of dynamical behaviors, including circular orbits and trapped stationary states. 
We identify the ratio of the phoretic mobilities between the two domains of the Janus particle as a central quantity in tuning their dynamics. 
Our results provide a path for developing optimum protocols for tuning the dynamics of phoretic Janus particles and mixing fluid at the microscale.
In addition, this work suggests a method for quantifying the surface properties of phoretic Janus particles, which have proven challenging to probe experimentally. 
\end{abstract}
\vspace{1.0cm}

Phoretic Janus particles are a class of colloids that show promise for applications related to mixing, sorting, and chemical delivery~\cite{xu2017light,venugopalan2020fantastic,zhang2021dual,zhang2021janus}. 
These typically micron-sized particles self-propel due to their ability to generate and sustain chemical gradients across their surface~\cite{anderson1989colloid,prieve1984motion}.
The particles' Janus nature, where their surface is composed of two chemically distinct regions, is a common design feature for introducing self-propulsion. 
Experimentally, it is possible to tune both the size and composition of these regions, and there now exists a sizeable catalog of phoretic Janus particles, including bimetallic and platinum-coated colloids, biodegradable Janus micromotors, and colloidal particles coated in two different enzymes~\cite{safdar2018progress,paxton2004catalytic,theurkauff2012dynamic,solovev2009catalytic,solovev2011light,gao2011highly,sanchez2011superfast,lee2014self,martin2015template,okmen2024multi,maiti2019self}.
An appropriate fuel source is a second design element nearly universal to these systems.
Hydrogen peroxide is a popular choice as one of the particle regions is usually metallic and catalytic. 
However, the fuel choice is flexible depending on the particle's composition.
For example, enzyme-coated particles use the corresponding substrate as the fuel source~\cite{Patino2018-in}. 
With their diverse compositions and fuel sources, phoretic Janus particles offer a robust design platform for engineering behavior at the microscale.

Notably, the ability of these particles to autonomously navigate complex microfluidic environments makes them an ideal candidate for a range of chemical delivery and sensing applications, including targeted drug delivery and environmental remediation~\cite{lu2015multifunctional,tan2023development,vilela2016graphene,villa2018metal,fu2023recent,soler2013self,wani2016dual,mushtaq2016highly,zhang2017light,wang2019photocatalytic,beladi2021maze,noauthor_undated-ca}.
For example, recent \textit{in situ} work shows CaCO$_3$ Janus particles exhibit a pH sensitivity that can induce a chemotactic response toward HeLa cancer cells~\cite{Guix2016-mp}.
Regarding environmental remediation, Wang et al.~\cite{wang2019photocatalytic} recently proposed a strategy for removing microplastics using photocatalytic TiO$_2$ Janus particles. 
Additionally, phoretic Janus can serve as a tool for fluid mixing and directing self-assembly at the microscale~\cite{Mallory2018-nn, Mallory2019-to}. 
Examples include restructuring colloidal gels by incorporating a small fraction of phoretic Janus particles into the gel network~\cite{Szakasits2017-hq, Szakasits2019-bx, Omar2019-xe, Mallory2020-cx}, powering primitive micromachines~\cite{Maggi2016-hb, Soto2022-qr, Liu2022-ku}, and generating bulk fluid flow by fabricating self-pumping walls patterned with Janus micropillar arrays~\cite{Yu2020-tm} or by trapping phoretic Janus particles near boundaries or surfaces~\cite{uspal2015self,bayati2019dynamics}.

Any application utilizing phoretic Janus particles requires a deep understanding of how they explore and respond to their environment. 
A ubiquitous feature of these systems is the role of gradients in chemical fuel concentration, which can dramatically affect their single particle and collective behavior~\cite{das2015boundaries,wang2015enhanced,simmchen2016topographical,liu2016bimetallic,wang2017janus,jalaal2022interfacial,yu2016confined,jiang2010active,auschra2021thermotaxis,chen2023steering,sharifi2016pair,mallory2017self,liebchen2018synthetic,stark2018artificial,sturmer2019chemotaxis,che2022light,Singh2024-ci}.  
Several studies have focused on the behavior of a single phoretic Janus particle in a linear fuel concentration gradient and shown phoretic Janus particles will undergo a chemotactic response where the particle will reorient to move parallel or anti-parallel to the gradient~\cite{saha2014clusters,popescuchemotaxis18,vinze2021motion,xiao2022platform}.
However, the characterization of the motion of phoretic Janus particles in chemical gradients of other geometries is limited, which is in {\color{black} stark contrast to the exploration of passive colloidal particles in external solute gradient, where several protocols have been developed to tune the dynamics and sort passive colloids based on their surface properties~\cite{Warren2020-vh, Raj2023-bx, Williams2024-xh}.}
For phoretic Janus particles, a critical case that has received lesser attention and the focus of this study is radially symmetric gradients generated from the presence of a point sink or source of the chemical fuel self-propelling the particle.

Using a combination of analytical theory and numerical simulation, we quantify the influence of the radial concentration field on the Janus particle's trajectory. 
We find the particle can exhibit a rich array of behaviors that strongly depend on its surface properties, initial configuration, and the strength of the sink or source. 
The particle can migrate toward or away from the sink or source, similar to a passive particle in an external chemical gradient.
In addition, we identify conditions that trap the particle in a stationary state at a fixed distance from the sink or source. 
Furthermore, the particle's motion is no longer rectilinear as in the case of a particle in a linear fuel concentration gradient but can undergo a spiraling motion.
We identify conditions that stabilize the spiraling trajectories leading to a circular orbit about the sink or source. 
As stationary and orbiting states offer an innovative way to blend and pump fluid continually, our findings suggest potential applications in fluid mixing and microscale transport. 
In addition, the characteristic dynamics of a phoretic Janus particle in the presence of a point sink or source can serve as a diagnostic tool to identify its surface properties, which has been an experimental challenge. 

\begin{figure}[b!]
\centering
\includegraphics[width=.475\textwidth]{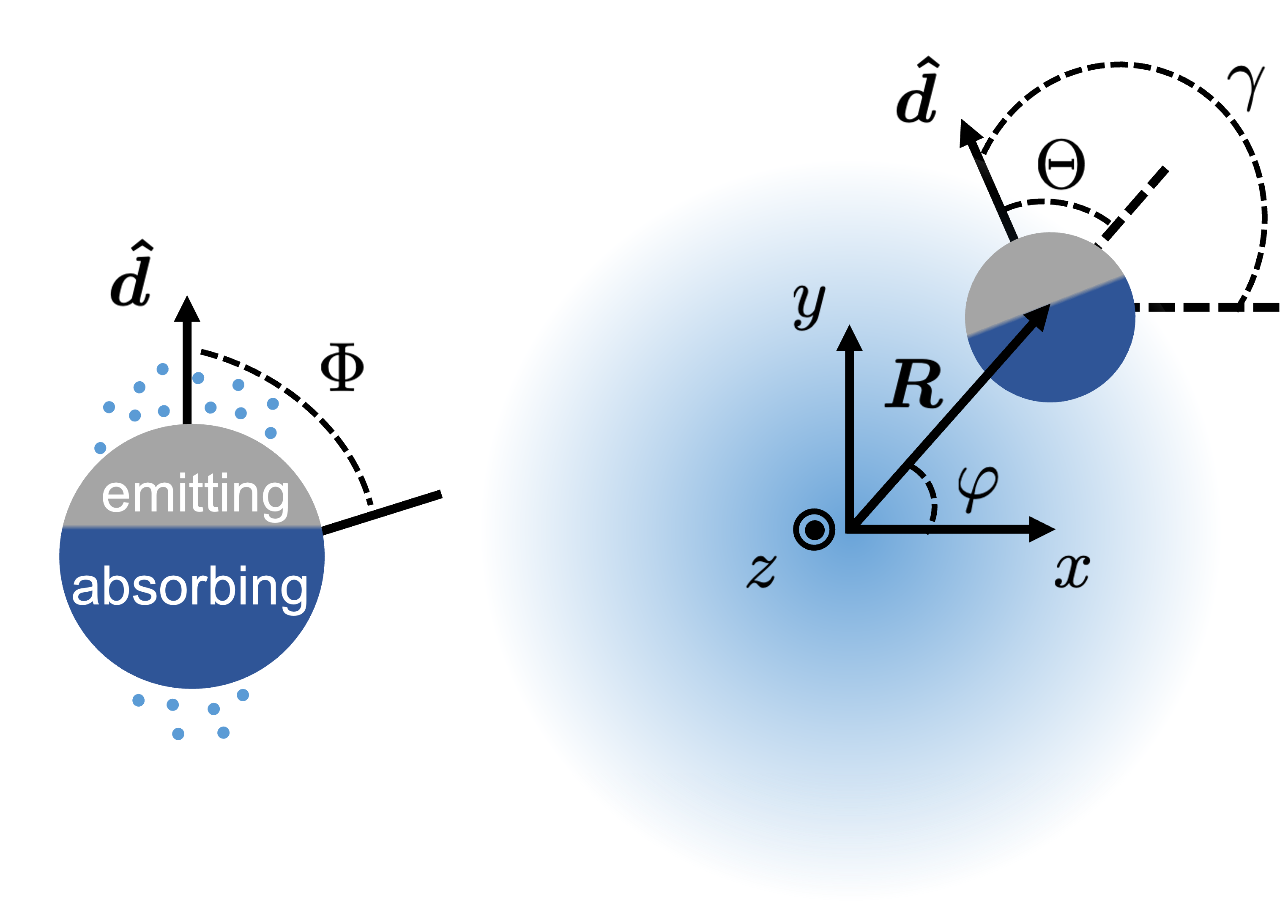}
\caption{\protect\small{Schematic of a spherical Janus particle with Janus balance $\chi = - \cos \Phi$ immersed in solution near a source or sink of the same chemical solute driving self-propulsion. The distance of the particle from the source $R$ and its orientation $\Theta = \cos^{-1} (\bm{\hat{R}} \cdot \bm{\hat{d}})$ fully specify the particle configuration.}}
\label{fig:figure_1} 
\end{figure}

\section{Results and Discussion}

We consider a phoretic Janus particle of diameter $a$ with bilateral symmetry along a predefined orientation unit vector $\bm{\hat{d}}$ as shown in Fig.~\ref{fig:figure_1}.
We implement a standard generic model for the phoretic self-propulsion mechanism where one region of the particle (gray) emits a particular chemical solute and the other region absorbs the solute (blue)~\cite{moran2017phoretic,bayati2016dynamics}. 
{\color{black} This model is amenable to describing the behavior of a broad range of phoretic Janus particles, including enzyme-coated Janus colloids, thermoresponsive Janus microgel particles, and bimetallic Janus colloids\cite{maiti2019self,schattling2015enhanced,hortelao2018targeting,ma2015enzyme,seo2013one,kurashina2021simultaneous,mou2014autonomous,paxton2004catalytic,dong2016highly}.}
The chemical solute represents the fuel driving self-propulsion.
The relative ratio of the absorbing to the emitting region is defined via the Janus balance $\chi$ such that a half-covered particle has $\chi = 0$, a particle emitting solute over its entire surface has $\chi = 1$, and for a fully absorbing particle $\chi = -1$.
To preserve mass balance, the rate of emission $Q_e$ and absorption $Q_a$ of the solute are constant, and there is no net change of solute such that $S_e Q_e - S_a Q_a = 0$, where $S_e$ and $S_a$ are the surface areas of the emitting and absorbing regions of the particle, respectively.
Under these steady-state conditions, a simple relationship exists between the emission and absorption rates of the two regions and the Janus balance $\chi$ given by $Q_a/Q_e = (1+\chi)/(1-\chi)$. 

Here, we focus on the athermal low Reynolds number limit, where Brownian motion is negligible, and particle motion is dominated solely by the phoretic self-propulsion mechanism.
{\color{black} For many phoretic Janus particles, neglecting the effects of Brownian motion is a reasonable approximation, as the self-propelling speed and diameter are sufficiently large that Brownian motion minimally influences their trajectories.}
In this regime, the particle's motion is deterministic and, as we demonstrate, confined to a two-dimensional plane.
The position of the particle relative to the source is given by a vector $\bm{R} = R\cos\varphi\, \bm{\hat{x}} + R\sin\varphi\, \bm{\hat{y}}$ where $R$ is the radial distance from the source and $\varphi$ is the angle between $\bm{R}$ and the positive $x$-axis as shown in Fig.~\ref{fig:figure_1}. 
We obtain the trajectory of the particle by integrating the equations of motion:
\begin{subequations}
    \label{eq:1}
    \begin{equation}
        \frac{dR}{dt} = U_R
        \label{subeq:1a}
    \end{equation}
    \begin{equation}
        \frac{d\varphi}{dt} = \frac{U_\varphi}{R}
        \label{subeq:1b}
    \end{equation}
     \begin{equation}
        \frac{d\gamma}{dt} =  \Omega_z
        \label{subeq:1c}
    \end{equation}     
\end{subequations}
\noindent where $U_R = \bm{U} \cdot \bm{\hat{R}}$ and $U_\varphi = \bm{U} \cdot \bm{\hat{\varphi}}$ are the radial and tangential components of the translation velocity $\bm{U}$, respectively.
The corresponding radial and tangential unit vectors are given by $\bm{\hat{R}} = \cos\varphi\, \bm{\hat{x}} + \sin\varphi\, \bm{\hat{y}}$ and $\bm{\hat{\varphi}} =  \bm{\hat{z}} \times \bm{\hat{R}} = -\sin\varphi\, \bm{\hat{x}} + \cos\varphi\,  \bm{\hat{y}}$.
The orientational dynamics of the particle are given by Eq.~(\ref{subeq:1c}) where $\gamma$ is the angle between the orientation vector $\bm{\hat{d}}$ and the positive $x$-axis and $\Omega_z$ is the only nonzero component of the angular velocity of the particle.
It is useful to define an auxiliary angle $\Theta = \gamma - \varphi$ which is the angle between $\bm{\hat{R}}$ and the orientation vector $\bm{\hat{d}}$, and from Eq.~(\ref{eq:1}b,c) it follows that $d\Theta/dt = d\gamma/dt - d\varphi/dt = \Omega_z - U_\varphi/R$.
Importantly, the configuration of the particle is fully determined by its orientation $\Theta$ and its distance $R$ from the singularity.

The Stokes equations prescribe the dynamics of the fluid and are given by
\begin{subequations}
    \label{eq:2}
    \begin{equation}
        \eta \nabla^2 \bm{u} - \nabla p = 0
        \label{subeq:2a}
    \end{equation}
    \begin{equation}
        \nabla \cdot \bm{u} = 0
        \label{subeq:2b}
    \end{equation}
\end{subequations}
\noindent where $\eta$, $\bm{u}$ and $p$ are the dynamic viscosity, fluid velocity, and pressure, respectively.
In the laboratory frame, the boundary conditions are $\bm{u} = 0$ at infinity and on the surface of the particle $\bm{u} = \bm{U} + \bm{\Omega} \times (\bm{r}-\bm{r}_0) + \bm{u}_s$ where $\bm{u}_{s}$ is the slip velocity of the fluid at the particle's surface.
Using the force and torque-free condition on the particle and the Lorentz reciprocal theorem~\cite{happel2012low}, the translation and angular velocity of a particle in an unbounded domain are~\cite{lauga2009hydrodynamics}
\begin{subequations}
    \label{eq:3}
    \begin{equation}
        \bm{U} = -\frac{1}{4 \pi a^2} \int_{s} \bm{u}_{s} ~ dS
        \label{subeq:3a}
    \end{equation}
    \begin{equation}
        \bm{\Omega} = -\frac{3}{8 \pi a^3} \int_{s} \bm{\hat{n}} \times \bm{u}_{s} ~  dS
        \label{subeq:3b}
    \end{equation}
\end{subequations}
where $dS$ is a differential element of the surface, and the domain of integration is over the entire surface of the particle $S$. 

Equation~(\ref{eq:3}) relates the slip velocity on the particle surface to the net motion of the particle.
The slip velocity arises from the interaction between the solute molecule and the particle's surface.
It is well-known that a solute gradient along the particle's surface induces an osmotic pressure gradient, generating a fluid flow within the Debye layer of the particle's surface.
In the thin Debye layer limit~\cite{anderson1989colloid}, this slip velocity is assumed to be located on the Janus particle surface and given by
\begin{equation}
    \bm{u}_{s} = - b(\bm{\hat{n}}) \nabla_{s} C\big|_{s}
    \label{eq:4}
\end{equation}
\noindent where $C$ is the fuel concentration field,~$\nabla_{s} = (\mathbb{I}-\bm{\hat{n}}\bm{\hat{n}}) \cdot \nabla$ the tangential projection of the surface gradient operator, and $\bm{\hat{n}}$ the normal unit vector on the Janus particle surface directed into the bulk solution.
The phoretic mobility $b(\bm{\hat{n}})$ can be either positive or negative and is determined by the details of the molecular interaction between the Janus particle and the solute particles~\cite{anderson1989colloid, Derjaguin1993-ph}.
We assume the particle's phoretic mobility is constant in a given surface region and denote the particle's absorbing and emitting sides as $b_a$ and $b_e$, respectively.
A central quantity in this study is the ratio of the phoretic mobilities of the two regions, which we call the phoretic mobility ratio and denote by $\beta = b_a/b_e$.

We treat the solute flux as being purely diffusive such that the fuel concentration field evolves according to the continuity equation $\partial_t C  =  D \nabla^2 C + \alpha ~ \delta(\bm{r})$ where $D$ is the diffusivity of the solute and $\alpha$ is the strength of the singularity, which can be positive or negative. 
Furthermore, it is reasonable to assume the fuel concentration field relaxes rapidly with respect to the particle's motion such that we can neglect its time dependence and assume $\partial C/\partial t = 0$.
Under these conditions, the fuel concentration $C$ is given by Laplace's equation with a point singularity at the origin
\begin{equation}
D \nabla^2 (C - C_{\infty}) + \alpha ~ \delta(\bm{r}) = 0 \, .
    \label{eq:5}
\end{equation}
The boundary condition on the surface of the Janus particle is given by
\begin{equation}
     - D \bm{\hat{n}} \cdot  \nabla C(\bm{r})  =  
 Q_{e} H(\bm{\hat{n}} \cdot \bm{\hat{d}} + \chi) - Q_{a} \left[1-H(\bm{\hat{n}} \cdot \bm{\hat{d}}+\chi)\right] 
 \label{eq:6}
\end{equation}
where $H(x)$ is the Heaviside function. The boundary condition for the fuel concentration is assumed to be constant at infinity and denoted by $C_{\infty}$. 

To summarize our workflow for obtaining the particle's trajectory, we first solve Eq.~(\ref{eq:5}) with the appropriate boundary conditions to determine the concentration gradient along the surface of the Janus particle. 
Once the gradient of the concentration field is known, we can compute the slip velocity along the surface of the particle via Eq.~(\ref{eq:4}), which in turn furnishes the translation and angular velocities via Eq.~(\ref{eq:3}).
The final step is to obtain the particle's trajectory by integrating Eq.~(\ref{eq:1}) with the known translational and angular velocities.

A central outcome of this work is an analytical solution for the translational and rotational velocity of the Janus particle as a function of the Janus balance $\chi$, ratio of the phoretic mobilities $\beta = b_a/b_e$, the characteristic velocity of the particle $U_0 = Q_eb_e/(2D)$, and the effective strength of the singularity $\Tilde{\alpha} = \alpha/(4\pi a^2 Q_e)$.
{\color{black} The characteristic velocity $U_0$ regulates the self-propelling speed of the particle and is a measure of the surface activity of the Janus particle relative to the diffusivity of the solute and can take on both positive and negative values depending on the sign of the phoretic mobility $b_e$. 
The effective strength of the source $\Tilde{\alpha}$ is a dimensionless quantity comparing the rate of solute production or absorption by the singularity to that of the Janus particle.}
For brevity, we present the final result for the Janus particle's velocity here, but a detailed derivation is available in the Supporting Information~\cite{note1}.
The radial, tangential, and angular velocities of the particle are given by
\begin{subequations}
    \label{eq:7}
      \begin{align}
       \label{subeq:7a}
    \frac{U_R}{U_0}
    & =  \Gamma [\chi, \beta] \cos \Theta
     - \Tilde{\alpha} \left(2\frac{a^2}{R^2} - (1 - \beta) M[\chi, R, \Theta]  \right)
      \end{align}
      \begin{align}
       \label{subeq:7b}
    \frac{U_\varphi}{U_0} & = \sin \Theta \bigg( \Gamma [\chi, \beta]  
    - \Tilde{\alpha} (1 - \beta) N[\chi, R, \Theta] \bigg)
    \end{align}
    \begin{equation}
        \frac{\Omega_z}{U_0/a}  =  - \frac{3}{2} \Tilde{\alpha}   (1 -\beta) \sin\Theta~ \omega[\chi, R, \Theta]
        \label{subeq:7c}
    \end{equation}
\end{subequations}
where $\Gamma [\chi, \beta]  = 1+\chi - (1 - \beta) B[\chi]$ is a positive dimensionless parameter that further modulates the particle's velocity. 
The functional form of the dimensionless Janus balance parameter $B[\chi]$ is given in the Supporting Information~\cite{note1}.
The dimensionless parameters $M[\chi, R, \Theta], N[\chi, R, \Theta]$, and $\omega[\chi, R, \Theta]$ are given as infinite series whose explicit forms are also include in the Supporting Information~\cite{note1}.
Each of these dimensionless parameters is a positive monotonically decreasing function of $R$ where the leading order term decays as $1/R^2$.
The analytical solution for the velocities and the resulting trajectories are in excellent agreement with Boundary Element Method simulations (see Supporting Information~\cite{note1}).

\begin{figure*}[t!]
	\centering
	\includegraphics[width=1.0\textwidth]{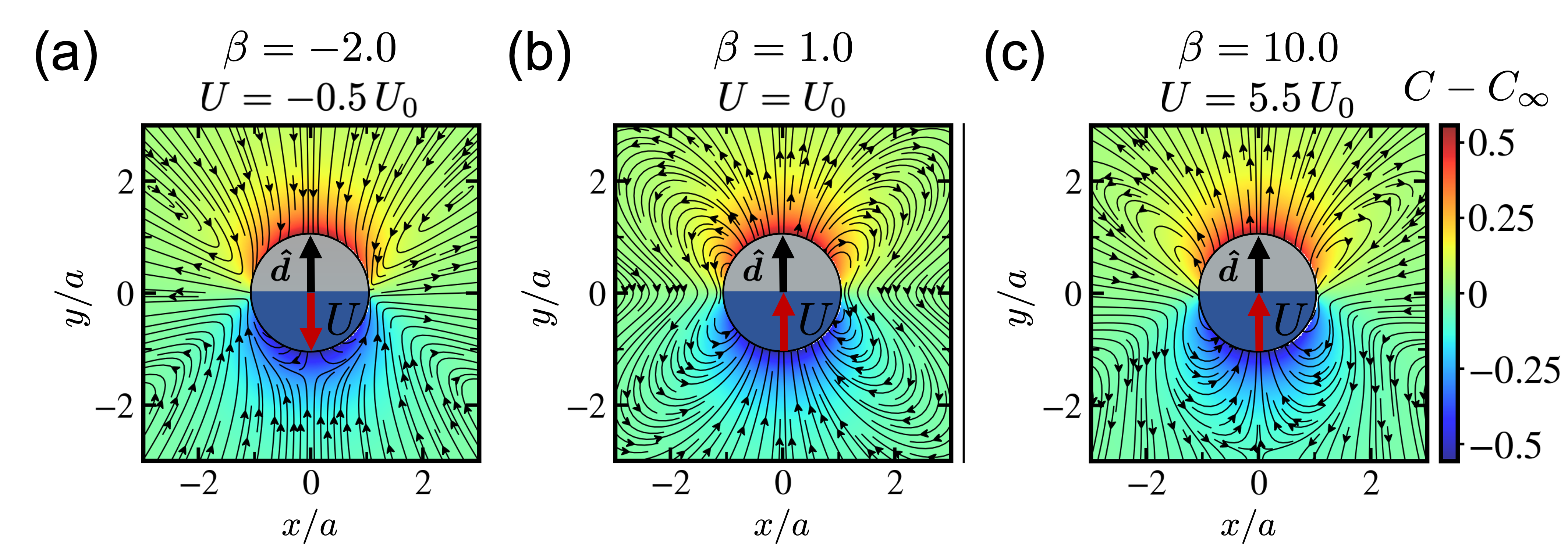}
	\caption{\protect\small{Local flow field in the laboratory frame and solute concentration gradient generated by a phoretic Janus particle in free space with phoretic mobility ratios (a) $\beta = -2.0$, (b) $\beta = 1.0$, and (c) $\beta = 10.0$. Here, we consider a particle with equal Janus balance $\chi = 0$ and characteristic velocity $U_0 = 1$.
 The red arrow indicates the direction of motion of the particle, where the grey and blue regions are the emitting and absorbing portions of the particle, respectively. The speed of the particle scales proportionally with the magnitude of the phoretic mobility ratio $|\beta|$. The flow field resembles a ``puller'' microswimmer when $\beta$ is negative and a ``pusher'' when positive. For $\beta = -1$ (not pictured), the velocity of the particle is zero as the flow field is antisymmetric about the orientation vector of the particle, resulting in no net motion.}}
	\label{fig:figure_2}
\end{figure*}

{\color{black} Prior to considering the role of the source or sink on the particle's behavior, we briefly review the motion of a phoretic Janus particle in free space by setting $\tilde{\alpha} = 0$ in Eq.~(\ref{eq:7}).
In this case, the velocity of the particle is constant, and the dimensionless parameter $\Gamma [\chi, \beta]$ and the characteristic velocity $U_0$ determine the particle's speed and whether its motion is parallel or antiparallel to the orientation vector $\bm{\hat{d}}$.
Figure~\ref{fig:figure_2} shows the direction of motion of a phoretic Janus particle in free space for various values of the phoretic mobility ratio $\beta$.
Here, we consider a particle with equal Janus balance $\chi = 0$ and characteristic velocity $U_0 = 1$.
The figure also includes the solute concentration field induced by the particle and highlights the changes in the flow field for different values of the phoretic mobility ratio $\beta$.
The solute concentration and flow fields were obtained by numerically solving Eq.~(\ref{eq:5}) and Eq.~(\ref{eq:2}), respectively, using BEM.
In general, the velocity of the particle vanishes for values of $\beta$ that satisfy
$\Gamma [\chi, \beta] = 1+\chi - (1 - \beta) B[\chi] = 0$, which occurs for $\beta = \, -1$ in the equal Janus balance case. 
For $\beta < -1$ [see Fig.~\ref{fig:figure_2}(a)], the particle moves with the absorbing side of the particle in front (i.e., along the $-\bm{\hat{d}}$ direction), and for $\beta > -1$, the particle moves with the emitting side of the particle in front [see Fig.~\ref{fig:figure_2}(b,c)]. 
Importantly, the concentration field induced by the particle is independent of the phoretic mobility ratio $\beta$, but the velocity and resulting flow field can vary widely.
Using the squirmer classification of microswimmers~\cite{Lighthill1952-um,Shen2018-jk}, particles with negative phoretic mobility ratio generate a local flow field that resembles a ``puller," while for positive values, the local flow field generated is more characteristic of a ``pusher".
Additionally, the speed of the particle $|U|$ increases as the magnitude of the phoretic mobility ratio increases $|\beta|$.
A phoretic Janus particle's velocity and the character of the resulting flow field are highly sensitive to the phoretic mobility ratio, which is what ultimately gives rise to the rich behavior in the presence of a radial concentration gradient. 
}

{\color{black} For a phoretic Janus particle in the presence of a point source or sink of solute}, the particle's motion is confined to the plane containing its orientation vector $\bm{\hat{d}}$ and radial vector $\bm{\hat{R}}$ and the only nonzero component of the angular velocity is normal to this plane.
{\color{black} As discussed previously}, the first terms of Eqs.~(\ref{subeq:7a}) and (\ref{subeq:7b}) correspond to the translational velocity of an isolated phoretic Janus particle. 
The second terms in each expression correspond to the velocity induced by the radial gradient generated by the singularity.
Far from the singularity (i.e., $R \rightarrow \infty$), these terms vanish, and Eq.~(\ref{eq:7}) reduces to that of a phoretic Janus particle in free space.
The angular velocity of the particle $\Omega_z$ [Eq.~(\ref{subeq:7c})] is due exclusively to the presence of the singularity and is strictly zero for $\beta = 1$, {\color{black} as the two phoretic mobilities are equal, and the slip velocity response is uniform across the surface of the particle}.
For all other values of $\beta${\color{black}, there is an asymmetric slip velocity response}, which can induce rotation.

As the motion of the Janus particle is deterministic and completely specified by its initial position and orientation, we find that unless there exists a fixed point, the particle will either eventually collide with the singularity or move off to infinity.
Thus, a natural scheme for classifying the dynamical behavior is identifying when fixed points occur as a function of the strength of the singularity $\tilde{\alpha}$ and the particle's surface properties. 
Importantly, our analytical solution for the particle's velocity [Eq.~(\ref{eq:7})] facilitates determining the location of fixed points and the resulting particle's trajectory for any $\tilde{\alpha}$, $\beta$, $\chi$, or $U_0$. 
However, for clarity, we restrict our discussion to a particle with equal Janus balance $\chi= 0$ and characteristic velocity $U_0 = 1$.
The equal Janus balance case is representative of many experimental systems and qualitatively illustrates the main features of the dynamics, including the emergence of fixed points.
In general, fixed points are a robust feature of a phoretic Janus particle in a radially symmetric gradient, and we observed qualitatively similar trends for particles with different Janus balances $\chi$.
The most critical parameters impacting a particle's dynamics are the phoretic mobility ratio $\beta$ and the strength of the singularity $\Tilde{\alpha}$.  
The strength of the singularity $\Tilde{\alpha}$ is the easiest parameter to tune experimentally, whereas the material properties of the Janus particle will determine the phoretic mobility ratio $\beta$.
Hence, we investigate the occurrence of fixed points as a function of $\tilde{\alpha}$ and $\beta$, which we refer to as the $\tilde{\alpha}\beta$-phase space.

{\color{black} For each point in the $\tilde{\alpha}\beta$-phase space, we conducted an exhaustive search as a function of $R$ and $\Theta$ to identify where the different components of the velocities vanish.
In our fixed point classification scheme, we recognize a \textit{trapped} or stationary state when $U_R$, $U_\varphi$ and $d \Theta / dt$ vanish and an orbiting state when only $U_R$ and $d \Theta / dt$ vanish with $U_{\varphi}$ remaining finite.
The defining feature of an orbiting state is that the particle's trajectory executes a closed circular path about the singularity.
By analyzing the eigenvalues of the Jacobian about a fixed point~\cite{Chasnov2022-gd}, we can further classify a fixed point as being stable, unstable, or a saddle point.
For stable fixed points, any small perturbation in $R$ or $\Theta$ will result in the particle returning to the initial fixed point, while unstable fixed points exhibit the opposite behavior. 
Saddle points are of a mixed character where it is stable for small perturbations in $R$ and unstable for perturbations in $\Theta$, or vice versa.
}

\begin{figure*}[t!]
	\centering
	\includegraphics[width=1.0\textwidth]{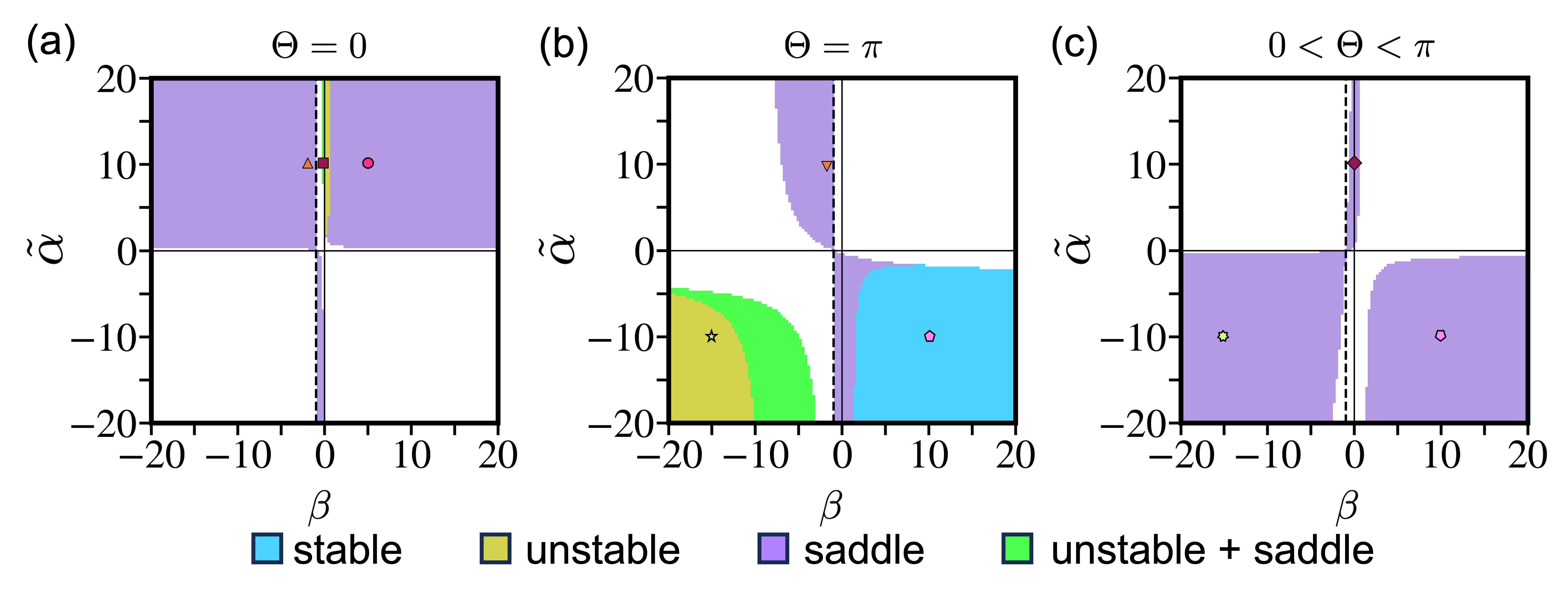}
	\caption{\protect\small{Classification of the different types of fixed points that can arise as a function of the phoretic mobility ratio $\beta$ and singularity strength $\tilde{\alpha}$ for a phoretic Janus particle with $\chi = 0$ and $U_0 = 1$. The fixed point state diagrams are organized based on the particle's orientation: (a) $\Theta = 0$, (b) $\Theta = \pi$, and  (c) $0 < \Theta < \pi$. The points correspond to representative fixed points in the $\tilde{\alpha}\beta$-phase space where the $R\Theta$-phase portrait and real-space trajectories are given in Fig.~\ref{fig:figure_4} and Fig.~\ref{fig:figure_5}. The dashed line at $\beta = -1$ corresponds to where the velocity vanishes for an isolated Janus particle, demarcating whether a particle will propel along or opposite its orientation vector.}}
	\label{fig:figure_3}
\end{figure*}

Importantly, as a function of the source strength $\tilde{\alpha}$, the occurrence and disappearance of a particular combination of fixed points can be used to deduce the approximate range of a particle's phoretic mobility ratio $\beta$ (see Fig.~\ref{fig:figure_3}). 
{\color{black} Due to the bilateral symmetry of the particle, it is easiest to locate fixed points when the orientation vector of the particle $\bm{\hat{d}}$ is either aligned or anti-aligned with the distance vector $\bm{\hat{R}}$.
For these orientations, $\Theta = 0$ and $\Theta = \pi$, both the tangential velocity $U_\varphi$ and the torque $\Omega_z$ on the particle are zero. 
From an experimental perspective, this may be the most straightforward approach to locating fixed points and estimating the phoretic mobility ratio.
If a particle can be fixed at either of these angles, then it only requires localizing it at various distances from the singularity and recording the response in the particle's motion.
In addition, these two orientations are the only ones where we find fixed points that are not saddle points. 
In Fig.~\ref{fig:figure_3}(a,b), we present the fixed point state diagram for particle orientations $\Theta = 0$ and $\Theta = \pi$ as a function of $\tilde{\alpha}$ and $\beta$.
For the values of $\tilde{\alpha}$ and $\beta$ located in the white regions of Fig.~\ref{fig:figure_3}, no fixed point exists for any position and orientation.
When the emitting side of the particle is furthest from the source ($\Theta = 0$), we find that most of the fixed points are stationary saddle points and occur predominately when $\tilde{\alpha} > 0$, except for a narrow region when $\beta \approx 1$ where there is a band of unstable stationary fixed points.} 

In the case that the absorbing side of the particle is further from the source ($\Theta = \pi$), the most experimentally relevant features are broad regions in the $\tilde{\alpha}\beta$-phase space where unstable and stable fixed points exist.
These unstable and stable fixed points occur when the particle is in the presence of a point sink ($\tilde{\alpha} < 0$), and will be discussed in greater detail in the next section.
{\color{black} In Fig.~\ref{fig:figure_3}(c), we highlight the fixed point state diagram for intermediate orientations between $0 < \Theta < \pi$ and find that all fixed points are orbiting saddle points.
Identifying fixed points for intermediate values of $\Theta$ presents a greater challenge as it requires identifying both values of $\Theta$ and $R$ where the velocities vanish. 
The different shaped points in Fig.~\ref{fig:figure_3} correspond to representative fixed points in the $\tilde{\alpha}\beta$-phase space, which will be investigated further in the next section.
At the end of the manuscript, we provide a more in-depth discussion of what ingredients are needed for the appearance of a fixed point, but the core requirement is the radial solute gradient generated by the source interacts with the particle in such a way that the velocity, and, more specifically, the integral of the slip velocity on the particle surface [Eq.~(\ref{eq:3})] vanishes.  
To meet this requirement, a careful balance between the concentration gradient about the surface of the particle and the phoretic mobility ratio needs to be satisfied.}

We now survey in more detail the different dynamical behaviors that arise by examining the $R\Theta$-phase portrait and the corresponding real space trajectories for selected values of $\tilde{\alpha}$ and $\beta$. 
For a known source strength $\tilde{\alpha}$, the location and type of the fixed point and, more generally, the $R\Theta$-phase portrait serve as a fingerprint for the particle's surface properties.  
\begin{figure*}[t!]
	\centering
        \includegraphics[width=1.\textwidth]{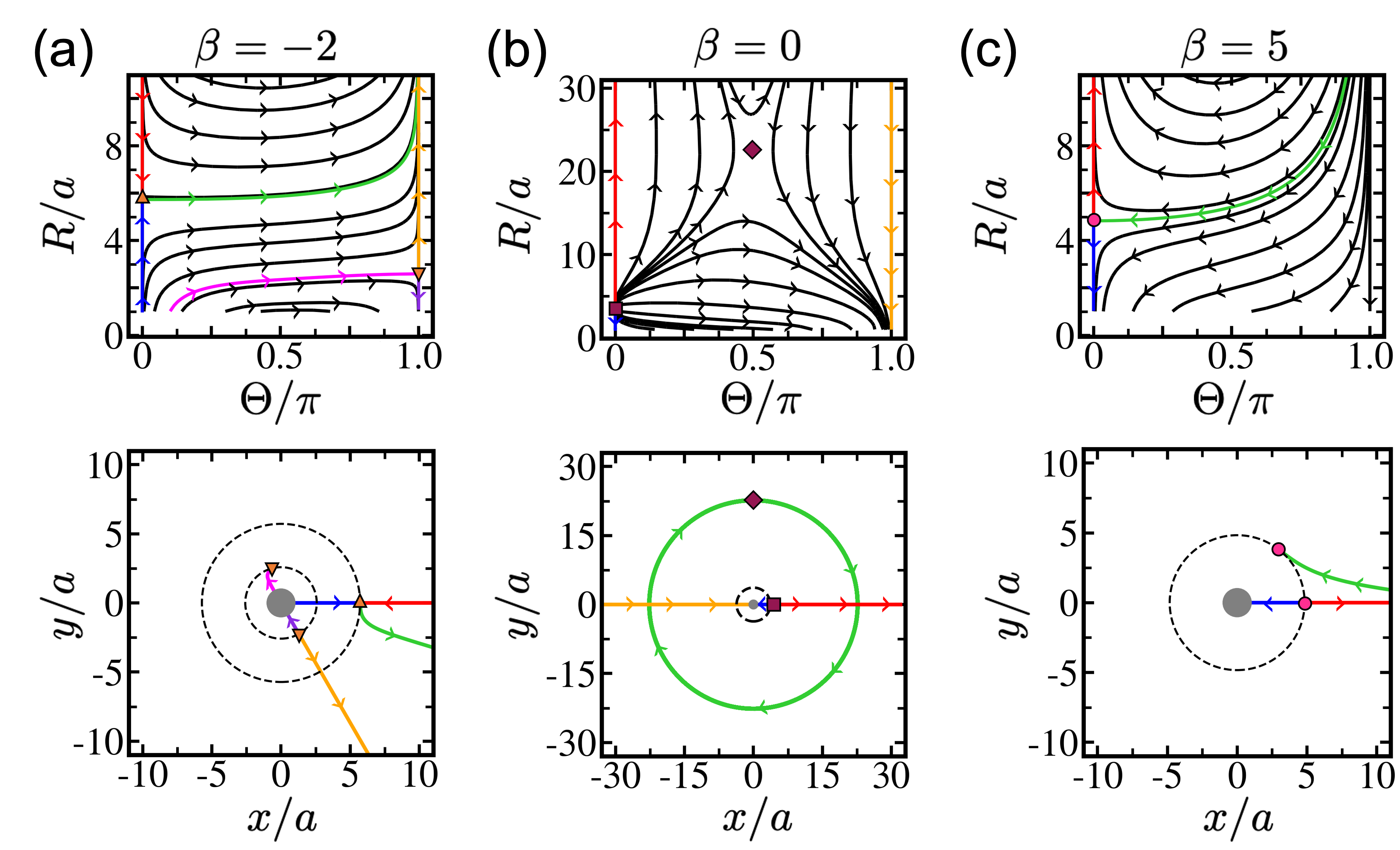}
	\caption{\protect\small{Dynamics near a chemical source: Phase portrait (top) and real-space trajectories (bottom) for phoretic Janus particles with different ratios of the phoretic mobilities $\beta$ in the presence of a point source $\Tilde{\alpha} = 10$. For each plot, particles have an equal Janus balance $\chi = 0$, and the characteristic velocity of the particle is $U_0 = 1$. The fixed points are indicated as points of various shapes, and the $R\Theta$-phase space trajectories and real space trajectories are color-coded accordingly. Videos of real-space trajectories are provided in the Supporting Information~\cite{note1}.}}
	\label{fig:figure_4}
\end{figure*}
{\color{black} To illustrate this connection, we first consider a representative source strength of $\tilde{\alpha} = 10$ and three different phoretic mobility ratios $\beta = -2,\, 0, \text{ and } 5$, which span the different fixed points regions outlined in the upper portion of Fig.~\ref{fig:figure_4}.
The $R\Theta$-phase portrait completely characterizes the motion of a phoretic Janus particle for a given choice of $\tilde{\alpha}$ and $\beta$, with each line corresponding to a specific trajectory where the initial position and orientation are located at a point on the line.
The arrows associated with each trajectory indicate the direction of motion. 
The bottom portion of Fig.~\ref{fig:figure_4} includes selected real-space trajectories for each of the $ R\Theta$-phase portraits and are color-coded accordingly. 
A comparison of the $R\Theta$-phase portraits highlights the sensitivity of the motion as a function of the phoretic mobility. 
The specific value of the phoretic mobility ratio will determine both the number of fixed points and their location. 
For the cases considered in Fig.~\ref{fig:figure_4}, all of the fixed points are stationary saddle points, with the exception of $\beta = 0$ [Fig.~\ref{fig:figure_4}(b)], where the farther fixed point is an orbiting saddle point, and the closer fixed point is an unstable stationary point. 
The orbiting saddle point would be a challenge to realize experimentally as it requires placing a particle at that specific distance and orientation from the source.
However, it would be possible to identify signatures of its existence by studying the trajectory of particles in the neighborhood of the orbiting fixed point.  
It is not only the location of the fixed points that will vary as a function of $\beta$, but also the direction of motion can be altered by the specific choice of the phoretic mobility ratio.
A representative example of this difference in behavior is to contrast the stationary saddle point at $\Theta = 0$ in Fig.~\ref{fig:figure_4}(a) and Fig.~\ref{fig:figure_4}(c). 
Both of these saddle points occur at a distance of $R/a \approx 5-6$, but the directions in which they are stable and unstable are reversed.}

\begin{figure*}[t!]
	\centering
        \includegraphics[width=1.\textwidth]{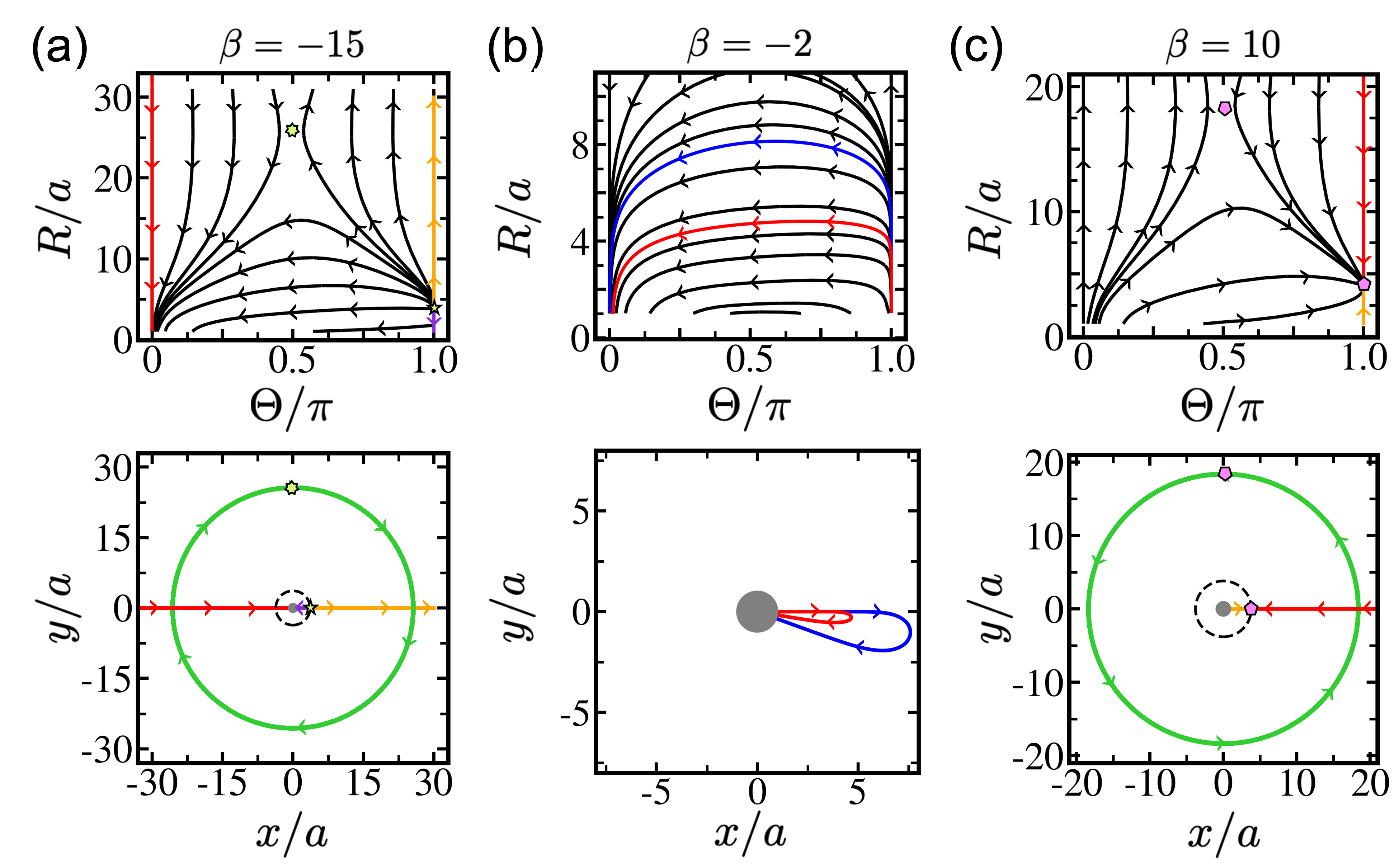}
	\caption{\protect\small{Dynamics near a chemical sink: $R\Theta$-phase portrait (top) and real-space trajectories (bottom) for phoretic Janus particles with different ratios of the phoretic mobilities $\beta$ in the presence of a point sink $\alpha = -10$. The other parameters are the same as Fig.~\ref{fig:figure_4} ($\chi = 0$ and $U_0 = 1$).  Videos of real-space trajectories are provided in the Supporting Information~\cite{note1}.
 }}
	\label{fig:figure_5}
\end{figure*}

In Figure~\ref{fig:figure_5}, we illustrate the different dynamical behaviors that can arise for a phoretic Janus particle in the presence of a point sink.
The strength of the singularity is $\tilde{\alpha} = -10$, which is equal in magnitude to that considered in Fig.~\ref{fig:figure_4}. 
We consider the behavior of particles with the three different phoretic mobility ratios $\beta = -15,\, -2, \text{ and } 10$.
In Fig.~\ref{fig:figure_5}(a), we find there are two fixed points, an orbiting saddle point and an unstable stationary point for $\beta = -15$. 
This unstable stationary point is within one of the few regions in the $\tilde{\alpha}\beta$-phase space where we observe an unstable fixed point. 
Interestingly, unstable fixed points will become stable under the reversal of the characteristic velocity $U_0$, {\color{black} which requires inverting the sign of the phoretic mobilities (i.e., $b_e \rightarrow -b_e$ and $b_a \rightarrow -b_a$)}
The sign of $U_0$ does not alter the location of the fixed points in the $\tilde{\alpha}\beta$-phase space but only leads to a reversal of the velocities. 
Thus, the $R\Theta$-phase portraits given in Fig.~\ref{fig:figure_4} and Fig.~\ref{fig:figure_5} for $U_0=1$ will have the same topology as $U_0 = -1$. 
However, the arrows indicating the direction of motion are reversed.

Figure~\ref{fig:figure_5}(b) is representative of a phoretic mobility ratio ($\beta = -2$) where there are no fixed points, and the particle eventually collides with the singularity for all orientations except $\Theta/\pi = 1$ where it moves off to infinity.
In Fig.~\ref{fig:figure_5}(c), we highlight the case of $\beta = 10$, which is representative of the only region of the $\tilde{\alpha}\beta$-phase space with a stable fixed point. 
A noteworthy feature of this region of the $\tilde{\alpha}\beta$-phase space is the $R\Theta$-phase portrait shows an exclusion region between the location of the singularity and the fixed point. 
Any particle initially in the region will either migrate toward the fixed point or off to infinity. 
This behavior mirrors the movement pattern observed in a recent study of \textit{P. aeruginosa} bacteria in response to a CO$_2$ point source, where the cells form an accumulation front at a distance from the source~\cite{shim2021co}.
Stable fixed points of this character have potential applications in surface cleaning from bio-contaminants or preventing the accumulation of active particles near surfaces.

\begin{figure}[t!]
	\centering
        \includegraphics[width=0.75\textwidth]{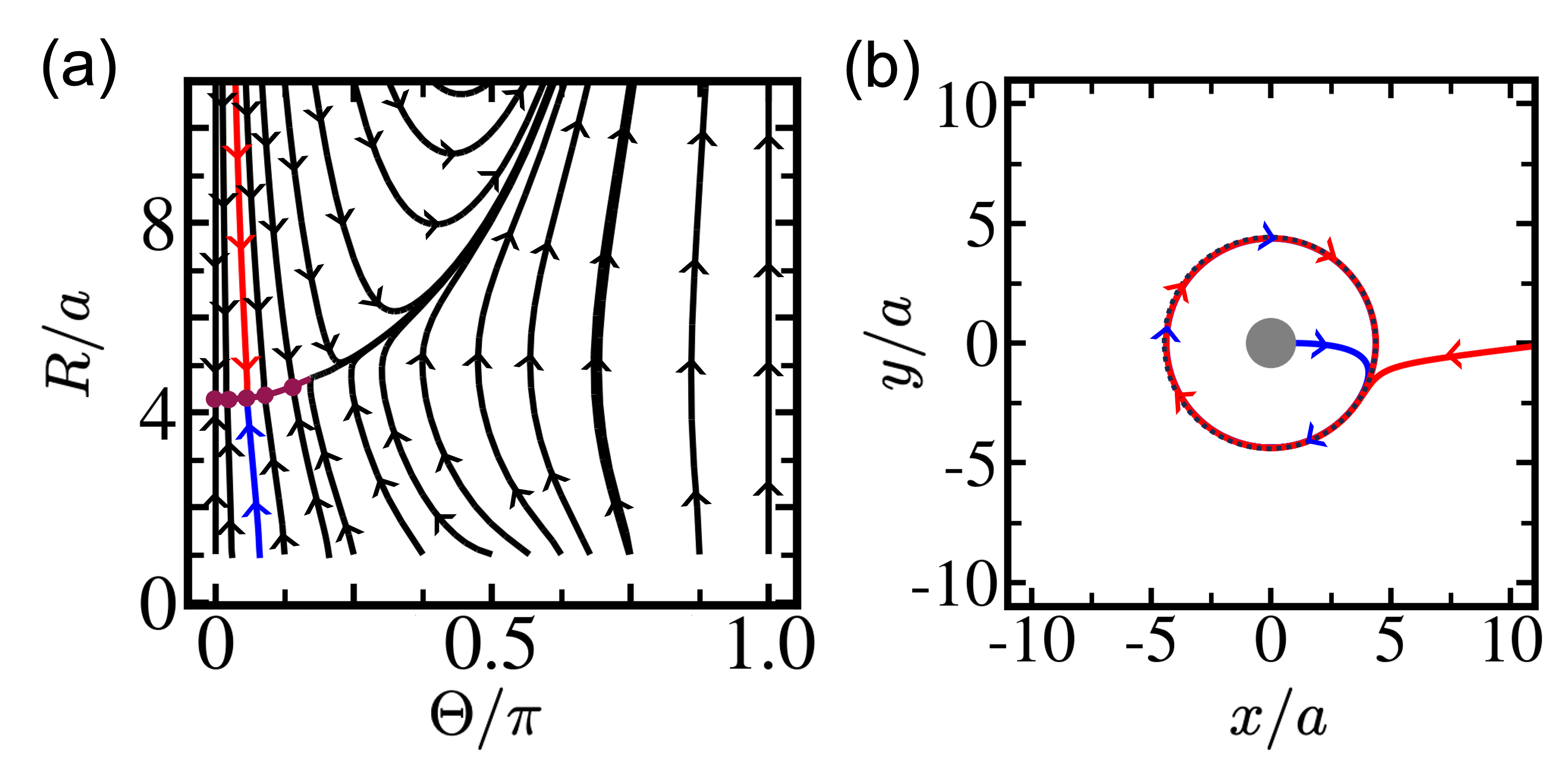}
	\caption{\protect\small{Infinite number of orbiting states: (a) $R\Theta$-phase portrait and (b) real-space trajectory in the region of the $\tilde{\alpha}\beta$-phase space where a continuous and infinite number of orbiting fixed points exists about the source. These plots correspond to the parameter values $\Tilde{\alpha} = 10$, $\beta = 0.644$, $\chi =0$ and $U_0 = -1$. Videos of the real-space trajectories of these orbiting states are provided in the Supporting Information~\cite{note1}. }}
 \label{fig:figure_6}
\end{figure}

We conclude our discussion of the $\tilde{\alpha}\beta$-phase space by highlighting a peculiar region not shown in Fig.~\ref{fig:figure_3},  where we observed a continuum of orbiting states.
A representative example of this behavior occurs for $U_0 = -1$, $\tilde{\alpha} = 10$ and $\beta \approx 0.64$, and the corresponding phase portrait and real space trajectories are given in Fig.~\ref{fig:figure_6}.
The magenta line in the $R\Theta$-phase portrait represents an infinite number of orbiting states that are stable with variations in $R$ and neutrally stable for $\Theta$.
Thus, any small changes in $\Theta$ or $R$ will lead to the particle finding a new orbit with a different radius.
This behavior, where there was observed to be a continuum of orbiting states, only occurs in a very small region of the $\tilde{\alpha}\beta$-phase space and will be further characterized in future work.
{\color{black} A Janus particle with this specific set of system parameters gives the highest likelihood of engineering a particle that will consistently undergo a circular orbit about the singularity.
This conclusion is inferred from the $R\Theta$-phase portrait, where there is a large region where the particle is drawn toward this continuum of orbiting fixed points, and these orbiting states are marginal stability in $\Theta$, making them more experimentally accessible than those discussed in Figs.~\ref{fig:figure_4} and~\ref{fig:figure_5}.}

Lastly, in the Supporting Information~\cite{note1}, we include figures that illustrate how the location of the fixed points change as a function of the strength of the singularity $\tilde{\alpha}$. 
The general trend is that the location of the fixed point moves further away from the singularity as its strength increases.
These trends agree with our physical interpretation, discussed in more detail in the next section. 
Essentially, when the strength of the singularity increases, so does the contribution to the particles' velocity from the presence of the singularity.
Thus, for a given characteristic velocity of the particle $U_0$, a fixed point will occur at a further distance from the singularity to compensate and appropriately cancel the increased contribution to the velocity from the singularity. 

\begin{figure*}[t!]
	\centering
        \includegraphics[width=0.75\textwidth]{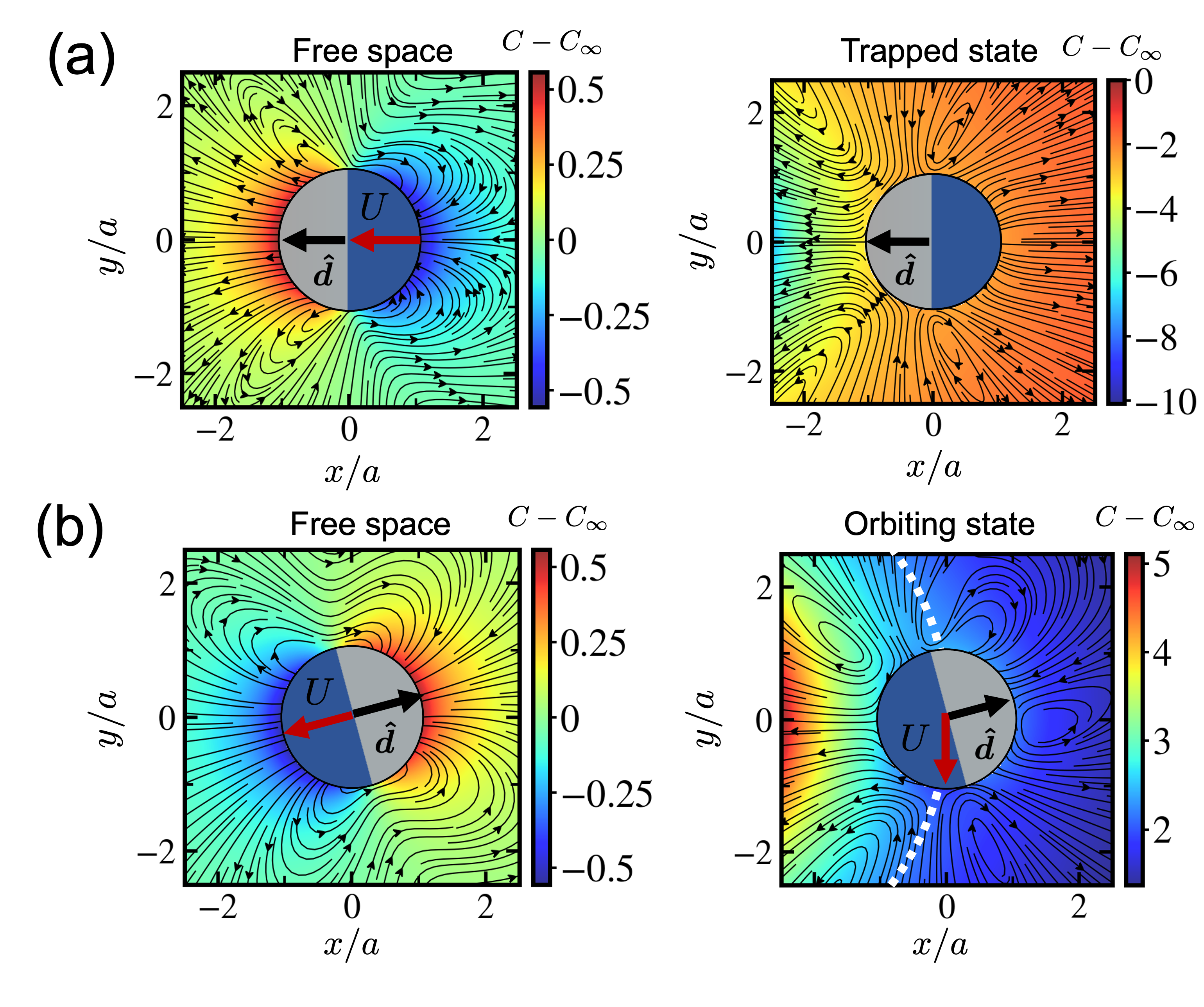}
	\caption{\protect\small{(a) Comparison between the flow field in the fixed laboratory frame and solute concentration field for a phoretic Janus particle with phoretic mobility ratio $\beta = 10$ in free space (left) and in the presence of a point sink of strength $\alpha = -10$ (right). At a distance of $R/a \approx 4$ and orientation $\Theta = \pi$, the particle is in a trapped stationary state as the velocity induced by the singularity cancels the characteristic velocity of the particle. (b) Comparison between the flow field and solute concentration field for a phoretic Janus particle with phoretic mobility ratio $\beta \approx 0.6$ in free space (left) and in the presence of a point source of strength $\alpha = 10$ (right). At a distance of $R/a \approx 4.4$ and orientation $\Theta \approx 0.086\pi$, the particle exists in an orbiting stationary state with tangential velocity $U_\varphi = -0.22 U_0$. The white dotted line illustrates the circular orbit the particle executes about the singularity.}}
	\label{fig:figure_7}
\end{figure*}

{\color{black} In this section, we provide a physical explanation for the existence of fixed points by exploring the link between the solute concentration gradient around the particle and the specific value of the phoretic mobility ratio.
The key factor is that the particle adopts an orientation and distance from the singularity such that the various components of its velocity are equal and opposite to the velocities induced by the presence of the singularity. 
In order for a fixed point to occur, the integral of the slip velocity along the particle's surface needs to vanish (Eq.~\ref{eq:3}), which requires a synergy between the concentration gradient a particle experiences at a certain distance and orientation from the singularity and the phoretic mobility ratio $\beta$.
In Fig.~\ref{fig:figure_7}, we plot the fuel concentration field and the resulting flow field for a stationary fixed point and an orbiting fixed point (along with the corresponding plot for the particle in free space). 
The stable stationary fixed point is due to the presence of a point sink ($\tilde{\alpha} = -10$), and the particle has a phoretic mobility ratio $\beta = 10$.
The $R\Theta$-phase portrait is that shown in Fig.~\ref{fig:figure_5}(c) where the fixed point occurs for $R/a \approx 4$ and $\Theta = \pi$. 
For the orientation $\Theta = \pi$, the tangential velocity $U_\varphi$ and the torque $\Omega_z$ are zero due to the bilateral symmetry of the particle. 
At a characteristic distance from the singularity, the particle's radial velocity will also vanish, which corresponds to the stationary state.
It is interesting to note that although the particle is stationary, it continues to pump fluid, as indicated by the resulting flow field.
In Fig.~\ref{fig:figure_7}(b), a similar plot is given comparing the flow and solute concentration field around an orbiting fixed state (corresponding to the $R\Theta$-phase portrait and trajectory plots shown in Fig.~\ref{fig:figure_6}) along with the corresponding plot for the particle in free space.
In this case, the particle is at a distance of $R/a \approx 4.4$ with an orientation $\Theta \approx 0.086\pi$  such that the radial component of the velocity and $d\Theta/dt$ vanish. 
The white dotted line illustrates the circular orbit the particle executes about the singularity.
It is a challenge to identify a general physical principle that can predict the stability criteria of a fixed point as it depends on the strength of the source $\tilde{\alpha}$, the phoretic mobility ratio $\beta$, the orientation $\Theta$ and distance from the singularity $R$.
Effectively, the stability of a fixed point relies on the particle's local response to the surrounding concentration field, and at present, we are unaware of a simple physical rationale for predicting their stability.
However, from our analytical solution, it is always possible to generate the $R\Theta$-phase portrait or compute the Jacobian, which provides stability criteria for any fixed point.  
} 

\section{Conclusions}

In this study, we quantified the dynamics of a phoretic Janus particle in a radially symmetric gradient generated by a point source or sink of the fuel driving self-propulsion. 
We derived an analytical expression for the phoretic Janus particle's velocity and found that its motion is highly sensitive to its surface properties and can exhibit various dynamical behaviors. 
In addition to positive and negative chemotaxis, we identify system parameters that give rise to circular orbits and trapped stationary states. 
We show that both types of fixed points are a robust feature of the $\tilde{\alpha}\beta$-phase space. 
The sensitivity of the location of the fixed points and, more generally, the topology of the $ R\Theta$-phase portrait that characterizes their trajectories suggests a method for quantifying the surface properties of phoretic Janus particles.

In addition, circular orbits and trapped stationary states offer a mechanism for pumping fluid and mixing at the microscale, particularly the stable stationary states, which are resistant to small fluctuations usually present in an experimental setting.
Even when a particle is trapped in a stationary or orbiting state, its surface is still chemically active and will pump fluid across its surface.
Our results demonstrate how to tune the location of a trapped or orbiting state via the strength of the source and the particle's phoretic mobility ratio.  
This ability to localize particles at a particular distance from the singularity suggests the possibility of achieving controlled fluid mixing at a desired rate, which is challenging at the microscale.

Future research related to this work includes investigating the role of chemical solute convection and the effect of Brownian motion on the dynamics in a radial chemical gradient.
{\color{black} Brownian motion can potentially randomize the trajectories of active particles if sufficiently strong, which would give rise to a new set of behaviors as the Janus particle's motion is no longer deterministic and can potentially lead to scenarios where a particle can be localized at multiple fixed points as it explores the phase portrait depending on its position and orientation.}
In addition, to better align the model to many experimental systems, we are currently investigating the role of a confining boundary, as many phoretic Janus particles are confined to move at a two-dimensional surface.

\section{Methods}

\textit{Simulation Details} - An explicit Euler scheme was implemented to integrate the particle's equations of motion [Eq.~(1)]. 
For a given initial position $R_0$ and orientation $\Theta_0$ of the particle, the trajectory is computed via 
\begin{subequations}
    \label{eq:D1}
    \begin{equation}
        R_{i+1} = R_i + U_R(R_i,\Theta_i) dt,
        \label{subeq:D1a}
    \end{equation}
    \begin{equation}
    \varphi_{i+1} = \varphi_i + \frac{U_\varphi(R_i,\Theta_i)}{R_i} dt,
        \label{subeq:C1b}
    \end{equation}
     \begin{equation}
     \gamma_{i+1} = \gamma_i + \Omega_z(R_i,\Theta_i) dt,
        \label{subeq:D1c}
    \end{equation}  
    \begin{equation}
     \Theta_{i} = \gamma_i - \varphi_i.
        \label{subeq:D1d}
    \end{equation}  
\end{subequations}
The velocities are calculated at each timestep via Eq.~(7) with the various infinite summations truncated at 400 terms. The Legendre polynomials were evaluated using the `scipy.special. eval\_legendre' function from the `scipy' Python library.
The accuracy of the analytical solution and its truncation were validated using Boundary Element Method (BEM) simulations to determine the particle velocity. 
This process involves solving the Laplace and Stokes equations in three dimensions, utilizing BEMLIB codes implemented in FORTRAN~\cite{pozrikidis2002practical} and specifically adapted for an active Janus particle~\cite{uspal2015self, bayati2016dynamics}. 
All BEM results agreed with the analytical solution [Eq.~(8)] and its truncated form.

\begin{acknowledgement}

The authors thank Igor Aronson, Ayusman Sen, Lauren Zarzar, and Will Noid for helpful discussion related to this work.
\end{acknowledgement}

\section{Associated Content}

A preprint of the submitted version of this manuscript is available on arXiv ~\cite{Bayati2024-px}.

\begin{suppinfo}

See Supporting Information at [URL] for complete derivation of the velocity of phoretic Janus particle, plot illustrating the radial dependence of dimensionless parameters in solution of Janus particles velocity, plot characterizing the location of fixed points as a function of the strength of the singularity, and movies illustrating the real space trajectories for different system parameters.

\end{suppinfo}


\begin{mcitethebibliography}{85}
\providecommand*\natexlab[1]{#1}
\providecommand*\mciteSetBstSublistMode[1]{}
\providecommand*\mciteSetBstMaxWidthForm[2]{}
\providecommand*\mciteBstWouldAddEndPuncttrue
  {\def\EndOfBibitem{\unskip.}}
\providecommand*\mciteBstWouldAddEndPunctfalse
  {\let\EndOfBibitem\relax}
\providecommand*\mciteSetBstMidEndSepPunct[3]{}
\providecommand*\mciteSetBstSublistLabelBeginEnd[3]{}
\providecommand*\EndOfBibitem{}
\mciteSetBstSublistMode{f}
\mciteSetBstMaxWidthForm{subitem}{(\alph{mcitesubitemcount})}
\mciteSetBstSublistLabelBeginEnd
  {\mcitemaxwidthsubitemform\space}
  {\relax}
  {\relax}

\bibitem[Xu \latin{et~al.}(2017)Xu, Mou, Gong, Luo, and Guan]{xu2017light}
Xu,~L.; Mou,~F.; Gong,~H.; Luo,~M.; Guan,~J. Light-Driven Micro/Nanomotors: From Fundamentals to Applications. \emph{Chem. Soc. Rev.} \textbf{2017}, \emph{46}, 6905--6926\relax
\mciteBstWouldAddEndPuncttrue
\mciteSetBstMidEndSepPunct{\mcitedefaultmidpunct}
{\mcitedefaultendpunct}{\mcitedefaultseppunct}\relax
\EndOfBibitem
\bibitem[Venugopalan \latin{et~al.}(2020)Venugopalan, Esteban-Fern{\'a}ndez~de {\'A}vila, Pal, Ghosh, and Wang]{venugopalan2020fantastic}
Venugopalan,~P.~L.; Esteban-Fern{\'a}ndez~de {\'A}vila,~B.; Pal,~M.; Ghosh,~A.; Wang,~J. Fantastic Voyage of Nanomotors Into the Cell. \emph{ACS Nano} \textbf{2020}, \emph{14}, 9423--9439\relax
\mciteBstWouldAddEndPuncttrue
\mciteSetBstMidEndSepPunct{\mcitedefaultmidpunct}
{\mcitedefaultendpunct}{\mcitedefaultseppunct}\relax
\EndOfBibitem
\bibitem[Zhang \latin{et~al.}(2021)Zhang, Li, Gao, Fan, Pang, Li, Wu, Xie, and He]{zhang2021dual}
Zhang,~H.; Li,~Z.; Gao,~C.; Fan,~X.; Pang,~Y.; Li,~T.; Wu,~Z.; Xie,~H.; He,~Q. Dual-Responsive Biohybrid Neutrobots for Active Target Delivery. \emph{Sci. Robot.} \textbf{2021}, \emph{6}, eaaz9519\relax
\mciteBstWouldAddEndPuncttrue
\mciteSetBstMidEndSepPunct{\mcitedefaultmidpunct}
{\mcitedefaultendpunct}{\mcitedefaultseppunct}\relax
\EndOfBibitem
\bibitem[Zhang \latin{et~al.}(2021)Zhang, Fu, Duan, Song, and Yang]{zhang2021janus}
Zhang,~X.; Fu,~Q.; Duan,~H.; Song,~J.; Yang,~H. {J}anus Nanoparticles: From Fabrication to (Bio) Applications. \emph{ACS Nano} \textbf{2021}, \emph{15}, 6147--6191\relax
\mciteBstWouldAddEndPuncttrue
\mciteSetBstMidEndSepPunct{\mcitedefaultmidpunct}
{\mcitedefaultendpunct}{\mcitedefaultseppunct}\relax
\EndOfBibitem
\bibitem[Anderson(1989)]{anderson1989colloid}
Anderson,~J.~L. Colloid Transport by Interfacial Forces. \emph{Annu. Rev. Fluid Mech.} \textbf{1989}, \emph{21}, 61--99\relax
\mciteBstWouldAddEndPuncttrue
\mciteSetBstMidEndSepPunct{\mcitedefaultmidpunct}
{\mcitedefaultendpunct}{\mcitedefaultseppunct}\relax
\EndOfBibitem
\bibitem[Prieve \latin{et~al.}(1984)Prieve, Anderson, Ebel, and Lowell]{prieve1984motion}
Prieve,~D.~C.; Anderson,~J.~L.; Ebel,~J.~P.; Lowell,~M.~E. Motion of a Particle Generated by Chemical Gradients. Part 2. Electrolytes. \emph{J. Fluid Mech.} \textbf{1984}, \emph{148}, 247--269\relax
\mciteBstWouldAddEndPuncttrue
\mciteSetBstMidEndSepPunct{\mcitedefaultmidpunct}
{\mcitedefaultendpunct}{\mcitedefaultseppunct}\relax
\EndOfBibitem
\bibitem[Safdar \latin{et~al.}(2018)Safdar, Khan, and J{\"a}nis]{safdar2018progress}
Safdar,~M.; Khan,~S.~U.; J{\"a}nis,~J. Progress Toward Catalytic Micro-and Nanomotors for Biomedical and Environmental Applications. \emph{Adv. Mater.} \textbf{2018}, \emph{30}, 1703660\relax
\mciteBstWouldAddEndPuncttrue
\mciteSetBstMidEndSepPunct{\mcitedefaultmidpunct}
{\mcitedefaultendpunct}{\mcitedefaultseppunct}\relax
\EndOfBibitem
\bibitem[Paxton \latin{et~al.}(2004)Paxton, Kistler, Olmeda, Sen, St.~Angelo, Cao, Mallouk, Lammert, and Crespi]{paxton2004catalytic}
Paxton,~W.~F.; Kistler,~K.~C.; Olmeda,~C.~C.; Sen,~A.; St.~Angelo,~S.~K.; Cao,~Y.; Mallouk,~T.~E.; Lammert,~P.~E.; Crespi,~V.~H. Catalytic Nanomotors: Autonomous Movement of Striped Nanorods. \emph{J. Am. Chem. Soc.} \textbf{2004}, \emph{126}, 13424--13431\relax
\mciteBstWouldAddEndPuncttrue
\mciteSetBstMidEndSepPunct{\mcitedefaultmidpunct}
{\mcitedefaultendpunct}{\mcitedefaultseppunct}\relax
\EndOfBibitem
\bibitem[Theurkauff \latin{et~al.}(2012)Theurkauff, Cottin-Bizonne, Palacci, Ybert, and Bocquet]{theurkauff2012dynamic}
Theurkauff,~I.; Cottin-Bizonne,~C.; Palacci,~J.; Ybert,~C.; Bocquet,~L. Dynamic Clustering in Active Colloidal Suspensions With Chemical Signaling. \emph{Phys. Rev. Lett.} \textbf{2012}, \emph{108}, 268303\relax
\mciteBstWouldAddEndPuncttrue
\mciteSetBstMidEndSepPunct{\mcitedefaultmidpunct}
{\mcitedefaultendpunct}{\mcitedefaultseppunct}\relax
\EndOfBibitem
\bibitem[Solovev \latin{et~al.}(2009)Solovev, Mei, Berm{\'u}dez~Ure{\~n}a, Huang, and Schmidt]{solovev2009catalytic}
Solovev,~A.~A.; Mei,~Y.; Berm{\'u}dez~Ure{\~n}a,~E.; Huang,~G.; Schmidt,~O.~G. Catalytic Microtubular Jet Engines Self-Propelled by Accumulated Gas Bubbles. \emph{Small} \textbf{2009}, \emph{5}, 1688--1692\relax
\mciteBstWouldAddEndPuncttrue
\mciteSetBstMidEndSepPunct{\mcitedefaultmidpunct}
{\mcitedefaultendpunct}{\mcitedefaultseppunct}\relax
\EndOfBibitem
\bibitem[Solovev \latin{et~al.}(2011)Solovev, Smith, Bof'Bufon, Sanchez, and Schmidt]{solovev2011light}
Solovev,~A.~A.; Smith,~E.~J.; Bof'Bufon,~C.~C.; Sanchez,~S.; Schmidt,~O.~G. Light-Controlled Propulsion of Catalytic Microengines. \emph{Angew. Chem.} \textbf{2011}, \emph{50}, 10875\relax
\mciteBstWouldAddEndPuncttrue
\mciteSetBstMidEndSepPunct{\mcitedefaultmidpunct}
{\mcitedefaultendpunct}{\mcitedefaultseppunct}\relax
\EndOfBibitem
\bibitem[Gao \latin{et~al.}(2011)Gao, Sattayasamitsathit, Orozco, and Wang]{gao2011highly}
Gao,~W.; Sattayasamitsathit,~S.; Orozco,~J.; Wang,~J. Highly Efficient Catalytic Microengines: Template Electrosynthesis of Polyaniline/Platinum Microtubes. \emph{J. Am. Chem. Soc.} \textbf{2011}, \emph{133}, 11862--11864\relax
\mciteBstWouldAddEndPuncttrue
\mciteSetBstMidEndSepPunct{\mcitedefaultmidpunct}
{\mcitedefaultendpunct}{\mcitedefaultseppunct}\relax
\EndOfBibitem
\bibitem[Sanchez \latin{et~al.}(2011)Sanchez, Ananth, Fomin, Viehrig, and Schmidt]{sanchez2011superfast}
Sanchez,~S.; Ananth,~A.~N.; Fomin,~V.~M.; Viehrig,~M.; Schmidt,~O.~G. Superfast Motion of Catalytic Microjet Engines at Physiological Temperature. \emph{J. Am. Chem. Soc.} \textbf{2011}, \emph{133}, 14860--14863\relax
\mciteBstWouldAddEndPuncttrue
\mciteSetBstMidEndSepPunct{\mcitedefaultmidpunct}
{\mcitedefaultendpunct}{\mcitedefaultseppunct}\relax
\EndOfBibitem
\bibitem[Lee \latin{et~al.}(2014)Lee, Alarcon-Correa, Miksch, Hahn, Gibbs, and Fischer]{lee2014self}
Lee,~T.-C.; Alarcon-Correa,~M.; Miksch,~C.; Hahn,~K.; Gibbs,~J.~G.; Fischer,~P. Self-Propelling Nanomotors in the Presence of Strong Brownian Forces. \emph{Nano Lett.} \textbf{2014}, \emph{14}, 2407--2412\relax
\mciteBstWouldAddEndPuncttrue
\mciteSetBstMidEndSepPunct{\mcitedefaultmidpunct}
{\mcitedefaultendpunct}{\mcitedefaultseppunct}\relax
\EndOfBibitem
\bibitem[Mart{\'\i}n \latin{et~al.}(2015)Mart{\'\i}n, Jurado-S{\'a}nchez, Escarpa, and Wang]{martin2015template}
Mart{\'\i}n,~A.; Jurado-S{\'a}nchez,~B.; Escarpa,~A.; Wang,~J. Template Electrosynthesis of High-Performance Graphene Microengines. \emph{Small} \textbf{2015}, \emph{11}, 3568--3574\relax
\mciteBstWouldAddEndPuncttrue
\mciteSetBstMidEndSepPunct{\mcitedefaultmidpunct}
{\mcitedefaultendpunct}{\mcitedefaultseppunct}\relax
\EndOfBibitem
\bibitem[Okmen~Altas \latin{et~al.}(2024)Okmen~Altas, Goktas, Topcu, and Aydogan]{okmen2024multi}
Okmen~Altas,~B.; Goktas,~C.; Topcu,~G.; Aydogan,~N. Multi-Stimuli-Responsive Tadpole-Like Polymer/Lipid {J}anus Microrobots for Advanced Smart Material Applications. \emph{ACS Appl. Mater. Interfaces.} \textbf{2024}, \relax
\mciteBstWouldAddEndPunctfalse
\mciteSetBstMidEndSepPunct{\mcitedefaultmidpunct}
{}{\mcitedefaultseppunct}\relax
\EndOfBibitem
\bibitem[Maiti \latin{et~al.}(2019)Maiti, Shklyaev, Balazs, and Sen]{maiti2019self}
Maiti,~S.; Shklyaev,~O.~E.; Balazs,~A.~C.; Sen,~A. Self-Organization of Fluids in a Multienzymatic Pump System. \emph{Langmuir} \textbf{2019}, \emph{35}, 3724--3732\relax
\mciteBstWouldAddEndPuncttrue
\mciteSetBstMidEndSepPunct{\mcitedefaultmidpunct}
{\mcitedefaultendpunct}{\mcitedefaultseppunct}\relax
\EndOfBibitem
\bibitem[Pati{\~n}o \latin{et~al.}(2018)Pati{\~n}o, Arqu{\'e}, Mestre, Palacios, and S{\'a}nchez]{Patino2018-in}
Pati{\~n}o,~T.; Arqu{\'e},~X.; Mestre,~R.; Palacios,~L.; S{\'a}nchez,~S. Fundamental Aspects of {Enzyme-Powered} Micro- And Nanoswimmers. \emph{Acc. Chem. Res.} \textbf{2018}, \emph{51}, 2662--2671\relax
\mciteBstWouldAddEndPuncttrue
\mciteSetBstMidEndSepPunct{\mcitedefaultmidpunct}
{\mcitedefaultendpunct}{\mcitedefaultseppunct}\relax
\EndOfBibitem
\bibitem[Lu \latin{et~al.}(2015)Lu, Liu, Li, Yu, Tang, Hu, and Ying]{lu2015multifunctional}
Lu,~C.; Liu,~X.; Li,~Y.; Yu,~F.; Tang,~L.; Hu,~Y.; Ying,~Y. Multifunctional {J}anus Hematite--Silica Nanoparticles: Mimicking Peroxidase-Like Activity and Sensitive Colorimetric Detection of Glucose. \emph{ACS Appl. Mater. Interfaces} \textbf{2015}, \emph{7}, 15395--15402\relax
\mciteBstWouldAddEndPuncttrue
\mciteSetBstMidEndSepPunct{\mcitedefaultmidpunct}
{\mcitedefaultendpunct}{\mcitedefaultseppunct}\relax
\EndOfBibitem
\bibitem[Tan \latin{et~al.}(2023)Tan, Danquah, Jeevanandam, and Barhoum]{tan2023development}
Tan,~K.~X.; Danquah,~M.~K.; Jeevanandam,~J.; Barhoum,~A. Development of {J}anus Particles as Potential Drug Delivery Systems for Diabetes Treatment and Antimicrobial Applications. \emph{Pharmaceutics} \textbf{2023}, \emph{15}, 423\relax
\mciteBstWouldAddEndPuncttrue
\mciteSetBstMidEndSepPunct{\mcitedefaultmidpunct}
{\mcitedefaultendpunct}{\mcitedefaultseppunct}\relax
\EndOfBibitem
\bibitem[Vilela \latin{et~al.}(2016)Vilela, Parmar, Zeng, Zhao, and S{\'a}nchez]{vilela2016graphene}
Vilela,~D.; Parmar,~J.; Zeng,~Y.; Zhao,~Y.; S{\'a}nchez,~S. Graphene-Based Microbots for Toxic Heavy Metal Removal and Recovery From Water. \emph{Nano Lett.} \textbf{2016}, \emph{16}, 2860--2866\relax
\mciteBstWouldAddEndPuncttrue
\mciteSetBstMidEndSepPunct{\mcitedefaultmidpunct}
{\mcitedefaultendpunct}{\mcitedefaultseppunct}\relax
\EndOfBibitem
\bibitem[Villa \latin{et~al.}(2018)Villa, Parmar, Vilela, and S{\'a}nchez]{villa2018metal}
Villa,~K.; Parmar,~J.; Vilela,~D.; S{\'a}nchez,~S. Metal-Oxide-Based Microjets for the Simultaneous Removal of Organic Pollutants and Heavy Metals. \emph{ACS Appl. Mater. Interfaces} \textbf{2018}, \emph{10}, 20478--20486\relax
\mciteBstWouldAddEndPuncttrue
\mciteSetBstMidEndSepPunct{\mcitedefaultmidpunct}
{\mcitedefaultendpunct}{\mcitedefaultseppunct}\relax
\EndOfBibitem
\bibitem[Fu \latin{et~al.}(2023)Fu, Zhang, Gao, Cui, Guan, Zhang, Zhang, and Peng]{fu2023recent}
Fu,~T.; Zhang,~B.; Gao,~X.; Cui,~S.; Guan,~C.-Y.; Zhang,~Y.; Zhang,~B.; Peng,~Y. Recent Progresses, Challenges, and Opportunities of Carbon-Based Materials Applied in Heavy Metal Polluted Soil Remediation. \emph{Sci. Total Environ.} \textbf{2023}, \emph{856}, 158810\relax
\mciteBstWouldAddEndPuncttrue
\mciteSetBstMidEndSepPunct{\mcitedefaultmidpunct}
{\mcitedefaultendpunct}{\mcitedefaultseppunct}\relax
\EndOfBibitem
\bibitem[Soler \latin{et~al.}(2013)Soler, Magdanz, Fomin, Sanchez, and Schmidt]{soler2013self}
Soler,~L.; Magdanz,~V.; Fomin,~V.~M.; Sanchez,~S.; Schmidt,~O.~G. Self-Propelled Micromotors for Cleaning Polluted Water. \emph{ACS Nano} \textbf{2013}, \emph{7}, 9611--9620\relax
\mciteBstWouldAddEndPuncttrue
\mciteSetBstMidEndSepPunct{\mcitedefaultmidpunct}
{\mcitedefaultendpunct}{\mcitedefaultseppunct}\relax
\EndOfBibitem
\bibitem[Wani \latin{et~al.}(2016)Wani, Safdar, Kinnunen, and J{\"a}nis]{wani2016dual}
Wani,~O.~M.; Safdar,~M.; Kinnunen,~N.; J{\"a}nis,~J. Dual Effect of Manganese Oxide Micromotors: Catalytic Degradation and Adsorptive Bubble Separation of Organic Pollutants. \emph{Chem. Eur. J.} \textbf{2016}, \emph{22}, 1244--1247\relax
\mciteBstWouldAddEndPuncttrue
\mciteSetBstMidEndSepPunct{\mcitedefaultmidpunct}
{\mcitedefaultendpunct}{\mcitedefaultseppunct}\relax
\EndOfBibitem
\bibitem[Mushtaq \latin{et~al.}(2016)Mushtaq, Asani, Hoop, Chen, Ahmed, Nelson, and Pan{\'e}]{mushtaq2016highly}
Mushtaq,~F.; Asani,~A.; Hoop,~M.; Chen,~X.-Z.; Ahmed,~D.; Nelson,~B.~J.; Pan{\'e},~S. Highly Efficient Coaxial TiO2-PtPd Tubular Nanomachines for Photocatalytic Water Purification With Multiple Locomotion Strategies. \emph{Adv. Funct. Mater.} \textbf{2016}, \emph{26}, 6995--7002\relax
\mciteBstWouldAddEndPuncttrue
\mciteSetBstMidEndSepPunct{\mcitedefaultmidpunct}
{\mcitedefaultendpunct}{\mcitedefaultseppunct}\relax
\EndOfBibitem
\bibitem[Zhang \latin{et~al.}(2017)Zhang, Dong, Wu, Gao, He, and Ren]{zhang2017light}
Zhang,~Q.; Dong,~R.; Wu,~Y.; Gao,~W.; He,~Z.; Ren,~B. Light-Driven {A}u-{Wo}3@{c} {J}anus Micromotors for Rapid Photodegradation of Dye Pollutants. \emph{ACS Appl. Mater. Interfaces} \textbf{2017}, \emph{9}, 4674--4683\relax
\mciteBstWouldAddEndPuncttrue
\mciteSetBstMidEndSepPunct{\mcitedefaultmidpunct}
{\mcitedefaultendpunct}{\mcitedefaultseppunct}\relax
\EndOfBibitem
\bibitem[Wang \latin{et~al.}(2019)Wang, Kaeppler, Fischer, and Simmchen]{wang2019photocatalytic}
Wang,~L.; Kaeppler,~A.; Fischer,~D.; Simmchen,~J. Photocatalytic TiO2 Micromotors for Removal of Microplastics and Suspended Matter. \emph{ACS Appl. Mater. Interfaces} \textbf{2019}, \emph{11}, 32937--32944\relax
\mciteBstWouldAddEndPuncttrue
\mciteSetBstMidEndSepPunct{\mcitedefaultmidpunct}
{\mcitedefaultendpunct}{\mcitedefaultseppunct}\relax
\EndOfBibitem
\bibitem[Beladi-Mousavi \latin{et~al.}(2021)Beladi-Mousavi, Hermanova, Ying, Plutnar, and Pumera]{beladi2021maze}
Beladi-Mousavi,~S.~M.; Hermanova,~S.; Ying,~Y.; Plutnar,~J.; Pumera,~M. A Maze in Plastic Wastes: Autonomous Motile Photocatalytic Microrobots Against Microplastics. \emph{ACS Appl. Mater. Interfaces} \textbf{2021}, \emph{13}, 25102--25110\relax
\mciteBstWouldAddEndPuncttrue
\mciteSetBstMidEndSepPunct{\mcitedefaultmidpunct}
{\mcitedefaultendpunct}{\mcitedefaultseppunct}\relax
\EndOfBibitem
\bibitem[Ghosh \latin{et~al.}(2020)Ghosh, Xu, Gupta, and Gracias]{noauthor_undated-ca}
Ghosh,~A.; Xu,~W.; Gupta,~N.; Gracias,~D.~H. Active Matter Therapeutics. \emph{Nano Today} \textbf{2020}, \emph{31}\relax
\mciteBstWouldAddEndPuncttrue
\mciteSetBstMidEndSepPunct{\mcitedefaultmidpunct}
{\mcitedefaultendpunct}{\mcitedefaultseppunct}\relax
\EndOfBibitem
\bibitem[Guix \latin{et~al.}(2016)Guix, Meyer, Koch, and Schmidt]{Guix2016-mp}
Guix,~M.; Meyer,~A.~K.; Koch,~B.; Schmidt,~O.~G. Carbonate-based Janus micromotors moving in ultra-light acidic environment generated by {HeLa} cells in situ. \emph{Sci. Rep.} \textbf{2016}, \emph{6}, 21701\relax
\mciteBstWouldAddEndPuncttrue
\mciteSetBstMidEndSepPunct{\mcitedefaultmidpunct}
{\mcitedefaultendpunct}{\mcitedefaultseppunct}\relax
\EndOfBibitem
\bibitem[Mallory \latin{et~al.}(2018)Mallory, Valeriani, and Cacciuto]{Mallory2018-nn}
Mallory,~S.~A.; Valeriani,~C.; Cacciuto,~A. An Active Approach to Colloidal {Self-Assembly}. \emph{Annu. Rev. Phys. Chem.} \textbf{2018}, \emph{69}, 59--79\relax
\mciteBstWouldAddEndPuncttrue
\mciteSetBstMidEndSepPunct{\mcitedefaultmidpunct}
{\mcitedefaultendpunct}{\mcitedefaultseppunct}\relax
\EndOfBibitem
\bibitem[Mallory and Cacciuto(2019)Mallory, and Cacciuto]{Mallory2019-to}
Mallory,~S.~A.; Cacciuto,~A. {Activity-Enhanced} {Self-Assembly} of a Colloidal Kagome Lattice. \emph{J. Am. Chem. Soc.} \textbf{2019}, \emph{141}, 2500--2507\relax
\mciteBstWouldAddEndPuncttrue
\mciteSetBstMidEndSepPunct{\mcitedefaultmidpunct}
{\mcitedefaultendpunct}{\mcitedefaultseppunct}\relax
\EndOfBibitem
\bibitem[Szakasits \latin{et~al.}(2017)Szakasits, Zhang, and Solomon]{Szakasits2017-hq}
Szakasits,~M.~E.; Zhang,~W.; Solomon,~M.~J. Dynamics of Fractal Cluster Gels With Embedded Active Colloids. \emph{Phys. Rev. Lett.} \textbf{2017}, \emph{119}, 058001\relax
\mciteBstWouldAddEndPuncttrue
\mciteSetBstMidEndSepPunct{\mcitedefaultmidpunct}
{\mcitedefaultendpunct}{\mcitedefaultseppunct}\relax
\EndOfBibitem
\bibitem[Szakasits \latin{et~al.}(2019)Szakasits, Saud, Mao, and Solomon]{Szakasits2019-bx}
Szakasits,~M.~E.; Saud,~K.~T.; Mao,~X.; Solomon,~M.~J. Rheological implications of embedded active matter in colloidal gels. \emph{Soft Matter} \textbf{2019}, \emph{15}, 8012--8021\relax
\mciteBstWouldAddEndPuncttrue
\mciteSetBstMidEndSepPunct{\mcitedefaultmidpunct}
{\mcitedefaultendpunct}{\mcitedefaultseppunct}\relax
\EndOfBibitem
\bibitem[Omar \latin{et~al.}(2019)Omar, Wu, Wang, and Brady]{Omar2019-xe}
Omar,~A.~K.; Wu,~Y.; Wang,~Z.-G.; Brady,~J.~F. Swimming to Stability: Structural and Dynamical Control via Active Doping. \emph{ACS Nano} \textbf{2019}, \emph{13}, 560--572\relax
\mciteBstWouldAddEndPuncttrue
\mciteSetBstMidEndSepPunct{\mcitedefaultmidpunct}
{\mcitedefaultendpunct}{\mcitedefaultseppunct}\relax
\EndOfBibitem
\bibitem[Mallory \latin{et~al.}(2020)Mallory, Bowers, and Cacciuto]{Mallory2020-cx}
Mallory,~S.~A.; Bowers,~M.~L.; Cacciuto,~A. Universal Reshaping of Arrested Colloidal Gels via Active Doping. \emph{J. Chem. Phys.} \textbf{2020}, \emph{153}, 084901\relax
\mciteBstWouldAddEndPuncttrue
\mciteSetBstMidEndSepPunct{\mcitedefaultmidpunct}
{\mcitedefaultendpunct}{\mcitedefaultseppunct}\relax
\EndOfBibitem
\bibitem[Maggi \latin{et~al.}(2016)Maggi, Simmchen, Saglimbeni, Katuri, Dipalo, De~Angelis, Sanchez, and Di~Leonardo]{Maggi2016-hb}
Maggi,~C.; Simmchen,~J.; Saglimbeni,~F.; Katuri,~J.; Dipalo,~M.; De~Angelis,~F.; Sanchez,~S.; Di~Leonardo,~R. {Self-Assembly} of Micromachining Systems Powered by Janus Micromotors. \emph{Small} \textbf{2016}, \emph{12}, 446--451\relax
\mciteBstWouldAddEndPuncttrue
\mciteSetBstMidEndSepPunct{\mcitedefaultmidpunct}
{\mcitedefaultendpunct}{\mcitedefaultseppunct}\relax
\EndOfBibitem
\bibitem[Soto \latin{et~al.}(2022)Soto, Karshalev, Zhang, Esteban Fernandez~de Avila, Nourhani, and Wang]{Soto2022-qr}
Soto,~F.; Karshalev,~E.; Zhang,~F.; Esteban Fernandez~de Avila,~B.; Nourhani,~A.; Wang,~J. Smart Materials for Microrobots. \emph{Chem. Rev.} \textbf{2022}, \emph{122}, 5365--5403\relax
\mciteBstWouldAddEndPuncttrue
\mciteSetBstMidEndSepPunct{\mcitedefaultmidpunct}
{\mcitedefaultendpunct}{\mcitedefaultseppunct}\relax
\EndOfBibitem
\bibitem[Liu \latin{et~al.}(2022)Liu, Xie, Price, Liu, He, and Kong]{Liu2022-ku}
Liu,~T.; Xie,~L.; Price,~C.-A.~H.; Liu,~J.; He,~Q.; Kong,~B. Controlled propulsion of micro/nanomotors: operational mechanisms, motion manipulation and potential biomedical applications. \emph{Chem. Soc. Rev.} \textbf{2022}, \emph{51}, 10083--10119\relax
\mciteBstWouldAddEndPuncttrue
\mciteSetBstMidEndSepPunct{\mcitedefaultmidpunct}
{\mcitedefaultendpunct}{\mcitedefaultseppunct}\relax
\EndOfBibitem
\bibitem[Yu \latin{et~al.}(2020)Yu, Athanassiadis, Popescu, Chikkadi, G{\"u}th, Singh, Qiu, and Fischer]{Yu2020-tm}
Yu,~T.; Athanassiadis,~A.~G.; Popescu,~M.~N.; Chikkadi,~V.; G{\"u}th,~A.; Singh,~D.~P.; Qiu,~T.; Fischer,~P. Microchannels With {Self-Pumping} Walls. \emph{ACS Nano} \textbf{2020}, \emph{14}, 13673--13680\relax
\mciteBstWouldAddEndPuncttrue
\mciteSetBstMidEndSepPunct{\mcitedefaultmidpunct}
{\mcitedefaultendpunct}{\mcitedefaultseppunct}\relax
\EndOfBibitem
\bibitem[Uspal \latin{et~al.}(2015)Uspal, Popescu, Dietrich, and Tasinkevych]{uspal2015self}
Uspal,~W.; Popescu,~M.~N.; Dietrich,~S.; Tasinkevych,~M. Self-Propulsion of a Catalytically Active Particle Near a Planar Wall: From Reflection to Sliding and Hovering. \emph{Soft Matter} \textbf{2015}, \emph{11}, 434--438\relax
\mciteBstWouldAddEndPuncttrue
\mciteSetBstMidEndSepPunct{\mcitedefaultmidpunct}
{\mcitedefaultendpunct}{\mcitedefaultseppunct}\relax
\EndOfBibitem
\bibitem[Bayati \latin{et~al.}(2019)Bayati, Popescu, Uspal, Dietrich, and Najafi]{bayati2019dynamics}
Bayati,~P.; Popescu,~M.~N.; Uspal,~W.~E.; Dietrich,~S.; Najafi,~A. Dynamics Near Planar Walls for Various Model Self-Phoretic Particles. \emph{Soft matter} \textbf{2019}, \emph{15}, 5644--5672\relax
\mciteBstWouldAddEndPuncttrue
\mciteSetBstMidEndSepPunct{\mcitedefaultmidpunct}
{\mcitedefaultendpunct}{\mcitedefaultseppunct}\relax
\EndOfBibitem
\bibitem[Das \latin{et~al.}(2015)Das, Garg, Campbell, Howse, Sen, Velegol, Golestanian, and Ebbens]{das2015boundaries}
Das,~S.; Garg,~A.; Campbell,~A.~I.; Howse,~J.; Sen,~A.; Velegol,~D.; Golestanian,~R.; Ebbens,~S.~J. Boundaries Can Steer Active {J}anus Spheres. \emph{Nat. {C}ommun.} \textbf{2015}, \emph{6}, 8999\relax
\mciteBstWouldAddEndPuncttrue
\mciteSetBstMidEndSepPunct{\mcitedefaultmidpunct}
{\mcitedefaultendpunct}{\mcitedefaultseppunct}\relax
\EndOfBibitem
\bibitem[Wang \latin{et~al.}(2015)Wang, In, Blanc, Nobili, and Stocco]{wang2015enhanced}
Wang,~X.; In,~M.; Blanc,~C.; Nobili,~M.; Stocco,~A. Enhanced Active Motion of {J}anus Colloids at the Water Surface. \emph{Soft Matter} \textbf{2015}, \emph{11}, 7376--7384\relax
\mciteBstWouldAddEndPuncttrue
\mciteSetBstMidEndSepPunct{\mcitedefaultmidpunct}
{\mcitedefaultendpunct}{\mcitedefaultseppunct}\relax
\EndOfBibitem
\bibitem[Simmchen \latin{et~al.}(2016)Simmchen, Katuri, Uspal, Popescu, Tasinkevych, and S{\'a}nchez]{simmchen2016topographical}
Simmchen,~J.; Katuri,~J.; Uspal,~W.~E.; Popescu,~M.~N.; Tasinkevych,~M.; S{\'a}nchez,~S. Topographical Pathways Guide Chemical Microswimmers. \emph{Nat. {C}ommun.} \textbf{2016}, \emph{7}, 1--9\relax
\mciteBstWouldAddEndPuncttrue
\mciteSetBstMidEndSepPunct{\mcitedefaultmidpunct}
{\mcitedefaultendpunct}{\mcitedefaultseppunct}\relax
\EndOfBibitem
\bibitem[Liu \latin{et~al.}(2016)Liu, Zhou, Wang, and Zhang]{liu2016bimetallic}
Liu,~C.; Zhou,~C.; Wang,~W.; Zhang,~H. Bimetallic Microswimmers Speed Up in Confining Channels. \emph{Phys. Rev. Lett.} \textbf{2016}, \emph{117}, 198001\relax
\mciteBstWouldAddEndPuncttrue
\mciteSetBstMidEndSepPunct{\mcitedefaultmidpunct}
{\mcitedefaultendpunct}{\mcitedefaultseppunct}\relax
\EndOfBibitem
\bibitem[Wang \latin{et~al.}(2017)Wang, In, Blanc, Wurger, Nobili, and Stocco]{wang2017janus}
Wang,~X.; In,~M.; Blanc,~C.; Wurger,~A.; Nobili,~M.; Stocco,~A. Janus Colloids Actively Rotating on the Surface of Water. \emph{Langmuir} \textbf{2017}, \emph{33}, 13766--13773\relax
\mciteBstWouldAddEndPuncttrue
\mciteSetBstMidEndSepPunct{\mcitedefaultmidpunct}
{\mcitedefaultendpunct}{\mcitedefaultseppunct}\relax
\EndOfBibitem
\bibitem[Jalaal \latin{et~al.}(2022)Jalaal, ten Hagen, Diddens, Lohse, and Marin]{jalaal2022interfacial}
Jalaal,~M.; ten Hagen,~B.; Diddens,~C.; Lohse,~D.; Marin,~A. Interfacial Aggregation of Self-Propelled {J}anus Colloids in Sessile Droplets. \emph{Phys. Rev. Fluids} \textbf{2022}, \emph{7}, 110514\relax
\mciteBstWouldAddEndPuncttrue
\mciteSetBstMidEndSepPunct{\mcitedefaultmidpunct}
{\mcitedefaultendpunct}{\mcitedefaultseppunct}\relax
\EndOfBibitem
\bibitem[Yu \latin{et~al.}(2016)Yu, Kopach, Misko, Vasylenko, Makarov, Marchesoni, Nori, Baraban, and Cuniberti]{yu2016confined}
Yu,~H.; Kopach,~A.; Misko,~V.~R.; Vasylenko,~A.~A.; Makarov,~D.; Marchesoni,~F.; Nori,~F.; Baraban,~L.; Cuniberti,~G. Confined Catalytic {J}anus Swimmers in a Crowded Channel: Geometry-Driven Rectification Transients and Directional Locking. \emph{Small} \textbf{2016}, \emph{12}, 5882--5890\relax
\mciteBstWouldAddEndPuncttrue
\mciteSetBstMidEndSepPunct{\mcitedefaultmidpunct}
{\mcitedefaultendpunct}{\mcitedefaultseppunct}\relax
\EndOfBibitem
\bibitem[Jiang \latin{et~al.}(2010)Jiang, Yoshinaga, and Sano]{jiang2010active}
Jiang,~H.-R.; Yoshinaga,~N.; Sano,~M. Active Motion of a {J}anus Particle by Self-Thermophoresis in a Defocused Laser Beam. \emph{Phys. Rev. Lett.} \textbf{2010}, \emph{105}, 268302\relax
\mciteBstWouldAddEndPuncttrue
\mciteSetBstMidEndSepPunct{\mcitedefaultmidpunct}
{\mcitedefaultendpunct}{\mcitedefaultseppunct}\relax
\EndOfBibitem
\bibitem[Auschra \latin{et~al.}(2021)Auschra, Bregulla, Kroy, and Cichos]{auschra2021thermotaxis}
Auschra,~S.; Bregulla,~A.; Kroy,~K.; Cichos,~F. Thermotaxis of {J}anus Particles. \emph{Eur. Phys. J. E} \textbf{2021}, \emph{44}, 90\relax
\mciteBstWouldAddEndPuncttrue
\mciteSetBstMidEndSepPunct{\mcitedefaultmidpunct}
{\mcitedefaultendpunct}{\mcitedefaultseppunct}\relax
\EndOfBibitem
\bibitem[Chen \latin{et~al.}(2023)Chen, Chen, Elsayed, Edwards, Liu, Peng, Zhang, Zhang, Wang, and Wheeler]{chen2023steering}
Chen,~X.; Chen,~X.; Elsayed,~M.; Edwards,~H.; Liu,~J.; Peng,~Y.; Zhang,~H.; Zhang,~S.; Wang,~W.; Wheeler,~A.~R. Steering Micromotors via Reprogrammable Optoelectronic Paths. \emph{ACS Nano} \textbf{2023}, \emph{17}, 5894--5904\relax
\mciteBstWouldAddEndPuncttrue
\mciteSetBstMidEndSepPunct{\mcitedefaultmidpunct}
{\mcitedefaultendpunct}{\mcitedefaultseppunct}\relax
\EndOfBibitem
\bibitem[Sharifi-Mood \latin{et~al.}(2016)Sharifi-Mood, Mozaffari, and C{\'o}rdova-Figueroa]{sharifi2016pair}
Sharifi-Mood,~N.; Mozaffari,~A.; C{\'o}rdova-Figueroa,~U.~M. Pair Interaction of Catalytically Active Colloids: From Assembly to Escape. \emph{J. Fluid Mech.} \textbf{2016}, \emph{798}, 910--954\relax
\mciteBstWouldAddEndPuncttrue
\mciteSetBstMidEndSepPunct{\mcitedefaultmidpunct}
{\mcitedefaultendpunct}{\mcitedefaultseppunct}\relax
\EndOfBibitem
\bibitem[Mallory \latin{et~al.}(2017)Mallory, Alarcon, Cacciuto, and Valeriani]{mallory2017self}
Mallory,~S.; Alarcon,~F.; Cacciuto,~A.; Valeriani,~C. Self-Assembly of Active Amphiphilic {J}anus Particles. \emph{New J. Phys.} \textbf{2017}, \emph{19}, 125014\relax
\mciteBstWouldAddEndPuncttrue
\mciteSetBstMidEndSepPunct{\mcitedefaultmidpunct}
{\mcitedefaultendpunct}{\mcitedefaultseppunct}\relax
\EndOfBibitem
\bibitem[Liebchen and Lowen(2018)Liebchen, and Lowen]{liebchen2018synthetic}
Liebchen,~B.; Lowen,~H. Synthetic Chemotaxis and Collective Behavior in Active Matter. \emph{Acc. Chem. Res.} \textbf{2018}, \emph{51}, 2982--2990\relax
\mciteBstWouldAddEndPuncttrue
\mciteSetBstMidEndSepPunct{\mcitedefaultmidpunct}
{\mcitedefaultendpunct}{\mcitedefaultseppunct}\relax
\EndOfBibitem
\bibitem[Stark(2018)]{stark2018artificial}
Stark,~H. Artificial Chemotaxis of Self-Phoretic Active Colloids: Collective Behavior. \emph{Acc. Chem. Res.} \textbf{2018}, \emph{51}, 2681--2688\relax
\mciteBstWouldAddEndPuncttrue
\mciteSetBstMidEndSepPunct{\mcitedefaultmidpunct}
{\mcitedefaultendpunct}{\mcitedefaultseppunct}\relax
\EndOfBibitem
\bibitem[St{\"u}rmer \latin{et~al.}(2019)St{\"u}rmer, Seyrich, and Stark]{sturmer2019chemotaxis}
St{\"u}rmer,~J.; Seyrich,~M.; Stark,~H. Chemotaxis in a Binary Mixture of Active and Passive Particles. \emph{J. Chem. Phys.} \textbf{2019}, \emph{150}\relax
\mciteBstWouldAddEndPuncttrue
\mciteSetBstMidEndSepPunct{\mcitedefaultmidpunct}
{\mcitedefaultendpunct}{\mcitedefaultseppunct}\relax
\EndOfBibitem
\bibitem[Che \latin{et~al.}(2022)Che, Zhang, Mou, Guo, Kauffman, Sen, and Guan]{che2022light}
Che,~S.; Zhang,~J.; Mou,~F.; Guo,~X.; Kauffman,~J.~E.; Sen,~A.; Guan,~J. Light-Programmable Assemblies of Isotropic Micromotors. \emph{Research} \textbf{2022}, \relax
\mciteBstWouldAddEndPunctfalse
\mciteSetBstMidEndSepPunct{\mcitedefaultmidpunct}
{}{\mcitedefaultseppunct}\relax
\EndOfBibitem
\bibitem[Singh \latin{et~al.}(2024)Singh, Raman, Tripathi, Sharma, Choudhary, and Mangal]{Singh2024-ci}
Singh,~K.; Raman,~H.; Tripathi,~S.; Sharma,~H.; Choudhary,~A.; Mangal,~R. Pair Interactions of {Self-Propelled} {SiO2-Pt} Janus Colloids: Chemically Mediated Encounters. \emph{Langmuir} \textbf{2024}, \relax
\mciteBstWouldAddEndPunctfalse
\mciteSetBstMidEndSepPunct{\mcitedefaultmidpunct}
{}{\mcitedefaultseppunct}\relax
\EndOfBibitem
\bibitem[Saha \latin{et~al.}(2014)Saha, Golestanian, and Ramaswamy]{saha2014clusters}
Saha,~S.; Golestanian,~R.; Ramaswamy,~S. Clusters, Asters, and Collective Oscillations in Chemotactic Colloids. \emph{Phys. Rev. E} \textbf{2014}, \emph{89}, 062316\relax
\mciteBstWouldAddEndPuncttrue
\mciteSetBstMidEndSepPunct{\mcitedefaultmidpunct}
{\mcitedefaultendpunct}{\mcitedefaultseppunct}\relax
\EndOfBibitem
\bibitem[Popescu \latin{et~al.}(2018)Popescu, Uspal, Bechinger, and Fischer]{popescuchemotaxis18}
Popescu,~M.~N.; Uspal,~W.~E.; Bechinger,~C.; Fischer,~P. Chemotaxis of Active {J}anus Nanoparticles. \emph{Nano Lett.} \textbf{2018}, \emph{18}, 5345--5349\relax
\mciteBstWouldAddEndPuncttrue
\mciteSetBstMidEndSepPunct{\mcitedefaultmidpunct}
{\mcitedefaultendpunct}{\mcitedefaultseppunct}\relax
\EndOfBibitem
\bibitem[Vinze \latin{et~al.}(2021)Vinze, Choudhary, and Pushpavanam]{vinze2021motion}
Vinze,~P.~M.; Choudhary,~A.; Pushpavanam,~S. Motion of an Active Particle in a Linear Concentration Gradient. \emph{Phys. Fluids} \textbf{2021}, \emph{33}, 032011\relax
\mciteBstWouldAddEndPuncttrue
\mciteSetBstMidEndSepPunct{\mcitedefaultmidpunct}
{\mcitedefaultendpunct}{\mcitedefaultseppunct}\relax
\EndOfBibitem
\bibitem[Xiao \latin{et~al.}(2022)Xiao, Nsamela, Garlan, and Simmchen]{xiao2022platform}
Xiao,~Z.; Nsamela,~A.; Garlan,~B.; Simmchen,~J. A Platform for Stop-Flow Gradient Generation to Investigate Chemotaxis. \emph{Angew. Chem.} \textbf{2022}, \emph{61}, e202117768\relax
\mciteBstWouldAddEndPuncttrue
\mciteSetBstMidEndSepPunct{\mcitedefaultmidpunct}
{\mcitedefaultendpunct}{\mcitedefaultseppunct}\relax
\EndOfBibitem
\bibitem[Warren(2020)]{Warren2020-vh}
Warren,~P.~B. {Non-Faradaic} Electric Currents in the {Nernst-Planck} Equations and Nonlocal Diffusiophoresis of Suspended Colloids in Crossed Salt Gradients. \emph{Phys. Rev. Lett.} \textbf{2020}, \emph{124}, 248004\relax
\mciteBstWouldAddEndPuncttrue
\mciteSetBstMidEndSepPunct{\mcitedefaultmidpunct}
{\mcitedefaultendpunct}{\mcitedefaultseppunct}\relax
\EndOfBibitem
\bibitem[Raj \latin{et~al.}(2023)Raj, Shields, and Gupta]{Raj2023-bx}
Raj,~R.~R.; Shields,~C.~W.,~4th; Gupta,~A. Two-dimensional diffusiophoretic colloidal banding: optimizing the spatial and temporal design of solute sinks and sources. \emph{Soft Matter} \textbf{2023}, \emph{19}, 892--904\relax
\mciteBstWouldAddEndPuncttrue
\mciteSetBstMidEndSepPunct{\mcitedefaultmidpunct}
{\mcitedefaultendpunct}{\mcitedefaultseppunct}\relax
\EndOfBibitem
\bibitem[Williams \latin{et~al.}(2024)Williams, Warren, Sear, and Keddie]{Williams2024-xh}
Williams,~I.; Warren,~P.~B.; Sear,~R.~P.; Keddie,~J.~L. Colloidal diffusiophoresis in crossed electrolyte gradients: Experimental demonstration of an ``action-at-a-distance'' effect predicted by the {Nernst-Planck} equations. \emph{Phys. Rev. Fluids} \textbf{2024}, \emph{9}, 014201\relax
\mciteBstWouldAddEndPuncttrue
\mciteSetBstMidEndSepPunct{\mcitedefaultmidpunct}
{\mcitedefaultendpunct}{\mcitedefaultseppunct}\relax
\EndOfBibitem
\bibitem[Moran and Posner(2017)Moran, and Posner]{moran2017phoretic}
Moran,~J.~L.; Posner,~J.~D. Phoretic Self-Propulsion. \emph{Annu. Rev. Fluid Mech.} \textbf{2017}, \emph{49}, 511--540\relax
\mciteBstWouldAddEndPuncttrue
\mciteSetBstMidEndSepPunct{\mcitedefaultmidpunct}
{\mcitedefaultendpunct}{\mcitedefaultseppunct}\relax
\EndOfBibitem
\bibitem[Bayati and Najafi(2016)Bayati, and Najafi]{bayati2016dynamics}
Bayati,~P.; Najafi,~A. Dynamics of Two Interacting Active {J}anus Particles. \emph{J. Chem. Phys.} \textbf{2016}, \emph{144}, 134901\relax
\mciteBstWouldAddEndPuncttrue
\mciteSetBstMidEndSepPunct{\mcitedefaultmidpunct}
{\mcitedefaultendpunct}{\mcitedefaultseppunct}\relax
\EndOfBibitem
\bibitem[Schattling \latin{et~al.}(2015)Schattling, Thingholm, and Stadler]{schattling2015enhanced}
Schattling,~P.; Thingholm,~B.; Stadler,~B. Enhanced diffusion of glucose-fueled Janus particles. \emph{Chem. Mater.} \textbf{2015}, \emph{27}, 7412--7418\relax
\mciteBstWouldAddEndPuncttrue
\mciteSetBstMidEndSepPunct{\mcitedefaultmidpunct}
{\mcitedefaultendpunct}{\mcitedefaultseppunct}\relax
\EndOfBibitem
\bibitem[Hortelao \latin{et~al.}(2018)Hortelao, Carrascosa, Murillo-Cremaes, Patino, and Sanchez]{hortelao2018targeting}
Hortelao,~A.~C.; Carrascosa,~R.; Murillo-Cremaes,~N.; Patino,~T.; Sanchez,~S. Targeting 3D bladder cancer spheroids with urease-powered nanomotors. \emph{ACS Nano} \textbf{2018}, \emph{13}, 429--439\relax
\mciteBstWouldAddEndPuncttrue
\mciteSetBstMidEndSepPunct{\mcitedefaultmidpunct}
{\mcitedefaultendpunct}{\mcitedefaultseppunct}\relax
\EndOfBibitem
\bibitem[Ma \latin{et~al.}(2015)Ma, Jannasch, Albrecht, Hahn, Miguel-L{\'o}pez, Schaffer, and S{\'a}nchez]{ma2015enzyme}
Ma,~X.; Jannasch,~A.; Albrecht,~U.-R.; Hahn,~K.; Miguel-L{\'o}pez,~A.; Schaffer,~E.; S{\'a}nchez,~S. Enzyme-powered hollow mesoporous Janus nanomotors. \emph{Nano Lett.} \textbf{2015}, \emph{15}, 7043--7050\relax
\mciteBstWouldAddEndPuncttrue
\mciteSetBstMidEndSepPunct{\mcitedefaultmidpunct}
{\mcitedefaultendpunct}{\mcitedefaultseppunct}\relax
\EndOfBibitem
\bibitem[Seo \latin{et~al.}(2013)Seo, Doh, and Kim]{seo2013one}
Seo,~K.~D.; Doh,~J.; Kim,~D.~S. One-step microfluidic synthesis of Janus microhydrogels with anisotropic thermo-responsive behavior and organophilic/hydrophilic loading capability. \emph{Langmuir} \textbf{2013}, \emph{29}, 15137--15141\relax
\mciteBstWouldAddEndPuncttrue
\mciteSetBstMidEndSepPunct{\mcitedefaultmidpunct}
{\mcitedefaultendpunct}{\mcitedefaultseppunct}\relax
\EndOfBibitem
\bibitem[Kurashina \latin{et~al.}(2021)Kurashina, Tsuchiya, Sakai, Maeda, Heo, Rossi, Choi, Yanagisawa, and Onoe]{kurashina2021simultaneous}
Kurashina,~Y.; Tsuchiya,~M.; Sakai,~A.; Maeda,~T.; Heo,~Y.~J.; Rossi,~F.; Choi,~N.; Yanagisawa,~M.; Onoe,~H. Simultaneous crosslinking induces macroscopically phase-separated microgel from a homogeneous mixture of multiple polymers. \emph{Appl. Mater. Today} \textbf{2021}, \emph{22}, 100937\relax
\mciteBstWouldAddEndPuncttrue
\mciteSetBstMidEndSepPunct{\mcitedefaultmidpunct}
{\mcitedefaultendpunct}{\mcitedefaultseppunct}\relax
\EndOfBibitem
\bibitem[Mou \latin{et~al.}(2014)Mou, Chen, Zhong, Yin, Ma, and Guan]{mou2014autonomous}
Mou,~F.; Chen,~C.; Zhong,~Q.; Yin,~Y.; Ma,~H.; Guan,~J. Autonomous Motion and Temperature-Controlled Drug Delivery of Mg/Pt-Poly (N-Isopropylacrylamide) {J}anus Micromotors Driven by Simulated Body Fluid and Blood Plasma. \emph{ACS Appl. Mater. Interfaces} \textbf{2014}, \emph{6}, 9897--9903\relax
\mciteBstWouldAddEndPuncttrue
\mciteSetBstMidEndSepPunct{\mcitedefaultmidpunct}
{\mcitedefaultendpunct}{\mcitedefaultseppunct}\relax
\EndOfBibitem
\bibitem[Dong \latin{et~al.}(2016)Dong, Zhang, Gao, Pei, and Ren]{dong2016highly}
Dong,~R.; Zhang,~Q.; Gao,~W.; Pei,~A.; Ren,~B. Highly efficient light-driven TiO2--Au Janus micromotors. \emph{ACS Nano} \textbf{2016}, \emph{10}, 839--844\relax
\mciteBstWouldAddEndPuncttrue
\mciteSetBstMidEndSepPunct{\mcitedefaultmidpunct}
{\mcitedefaultendpunct}{\mcitedefaultseppunct}\relax
\EndOfBibitem
\bibitem[Happel and Brenner(2012)Happel, and Brenner]{happel2012low}
Happel,~J.; Brenner,~H. \emph{Low Reynolds Number Hydrodynamics: With Special Applications to Particulate Media}; Springer, 2012\relax
\mciteBstWouldAddEndPuncttrue
\mciteSetBstMidEndSepPunct{\mcitedefaultmidpunct}
{\mcitedefaultendpunct}{\mcitedefaultseppunct}\relax
\EndOfBibitem
\bibitem[Lauga and Powers(2009)Lauga, and Powers]{lauga2009hydrodynamics}
Lauga,~E.; Powers,~T.~R. The Hydrodynamics of Swimming Microorganisms. \emph{Rep. Prog. Phys.} \textbf{2009}, \emph{72}, 096601\relax
\mciteBstWouldAddEndPuncttrue
\mciteSetBstMidEndSepPunct{\mcitedefaultmidpunct}
{\mcitedefaultendpunct}{\mcitedefaultseppunct}\relax
\EndOfBibitem
\bibitem[Derjaguin \latin{et~al.}(1993)Derjaguin, Sidorenkov, Zubashchenko, and Kiseleva]{Derjaguin1993-ph}
Derjaguin,~B.~V.; Sidorenkov,~G.; Zubashchenko,~E.; Kiseleva,~E. Kinetic Phenomena in the Boundary Layers of Liquids 1. The Capillary Osmosis. \emph{Prog. Surf. Sci.} \textbf{1993}, \emph{43}, 138--152\relax
\mciteBstWouldAddEndPuncttrue
\mciteSetBstMidEndSepPunct{\mcitedefaultmidpunct}
{\mcitedefaultendpunct}{\mcitedefaultseppunct}\relax
\EndOfBibitem
\bibitem[not()]{note1}
See Supporting Information at [URL] for complete derivation of the velocity of phoretic Janus particle, plot illustrating the radial dependence of dimensionless parameters in solution of Janus particles velocity, plot characterizing the location of fixed points as a function of the strength of the singularity, and movies illustrating the real space trajectories for different system parameters.\relax
\mciteBstWouldAddEndPunctfalse
\mciteSetBstMidEndSepPunct{\mcitedefaultmidpunct}
{}{\mcitedefaultseppunct}\relax
\EndOfBibitem
\bibitem[Lighthill(1952)]{Lighthill1952-um}
Lighthill,~M.~J. On the squirming motion of nearly spherical deformable bodies through liquids at very small reynolds numbers. \emph{Commun. Pure Appl. Math.} \textbf{1952}, \emph{5}, 109--118\relax
\mciteBstWouldAddEndPuncttrue
\mciteSetBstMidEndSepPunct{\mcitedefaultmidpunct}
{\mcitedefaultendpunct}{\mcitedefaultseppunct}\relax
\EndOfBibitem
\bibitem[Shen \latin{et~al.}(2018)Shen, W{\"u}rger, and Lintuvuori]{Shen2018-jk}
Shen,~Z.; W{\"u}rger,~A.; Lintuvuori,~J.~S. Hydrodynamic interaction of a self-propelling particle with a wall. \emph{Eur. Phys. J. E Soft Matter} \textbf{2018}, \emph{41}, 39\relax
\mciteBstWouldAddEndPuncttrue
\mciteSetBstMidEndSepPunct{\mcitedefaultmidpunct}
{\mcitedefaultendpunct}{\mcitedefaultseppunct}\relax
\EndOfBibitem
\bibitem[Chasnov(2022)]{Chasnov2022-gd}
Chasnov,~J.~R. \emph{Differential Equations for Engineers (Mathematics for Engineers)}; Independently published, 2022\relax
\mciteBstWouldAddEndPuncttrue
\mciteSetBstMidEndSepPunct{\mcitedefaultmidpunct}
{\mcitedefaultendpunct}{\mcitedefaultseppunct}\relax
\EndOfBibitem
\bibitem[Shim \latin{et~al.}(2021)Shim, Khodaparast, Lai, Yan, Ault, Rallabandi, Shardt, and Stone]{shim2021co}
Shim,~S.; Khodaparast,~S.; Lai,~C.-Y.; Yan,~J.; Ault,~J.~T.; Rallabandi,~B.; Shardt,~O.; Stone,~H.~A. CO2-Driven Diffusiophoresis for Maintaining a Bacteria-Free Surface. \emph{Soft Matter} \textbf{2021}, \emph{17}, 2568--2576\relax
\mciteBstWouldAddEndPuncttrue
\mciteSetBstMidEndSepPunct{\mcitedefaultmidpunct}
{\mcitedefaultendpunct}{\mcitedefaultseppunct}\relax
\EndOfBibitem
\end{mcitethebibliography}

\providecommand{\noopsort}[1]{}\providecommand{\singleletter}[1]{#1}
\providecommand{\latin}[1]{#1}
\makeatletter
\providecommand{\doi}
  {\begingroup\let\do\@makeother\dospecials
  \catcode`\{=1 \catcode`\}=2 \doi@aux}
\providecommand{\doi@aux}[1]{\endgroup\texttt{#1}}
\makeatother
\providecommand*\mcitethebibliography{\thebibliography}
\csname @ifundefined\endcsname{endmcitethebibliography}  {\let\endmcitethebibliography\endthebibliography}{}

\end{document}


\subsubsection{Analytical solution for phoretic Janus particle velocity}
Due to the linearity of our governing equations [Eqs. (2) and (5) of the main text], we can apply the principle of superposition and decompose the 
solution for the total fuel solute concentration field into a sum of three elementary solutions, which we write as 
\begin{equation}
    C(\bm{r})-C_\infty = C_{sin}(\bm{r}) + C_{a}(\bm{r}) + C_d(\bm{r}) \, , 
        \label{eq:A1}
\end{equation}
where $C_\infty$ is the constant bulk fuel concentration far from the particle and singularity.
The concentration field $C_{sin}$ corresponds to the problem of a point singularity of strength $\alpha$ centered at the origin $\bm{r} = 0$, whose behavior is governed by
\begin{equation}
    D \nabla^2 C_{sin} + \alpha ~ \delta(\bm{r}) = 0 \, , 
        \label{eq:A2}
\end{equation}
with far-field boundary condition $C_{sin}|_{r\rightarrow \infty} = 0$.
The concentration field $C_{a}$ corresponds to the problem of an isolated phoretic Janus particle with governing equation
\begin{equation}
    \nabla^2 C_{a} = 0 \, , 
    \label{eq:A3}
\end{equation}
with the boundary condition on the surface of the particle given by Eq.~(6) of the main text and far-field boundary condition $C_{a}|_{r\rightarrow \infty} = 0$.
The concentration field $C_{d}$ also satisfies Laplace's equation, and is introduced to ensure the intrinsic activity of the Janus particle is not altered by the presence of the source.
The requisite boundary condition along the surface of the particle is
\begin{equation}
    \bm{\hat{n}} \cdot \nabla C_{d}|_{s}  = - \bm{\hat{n}} \cdot \nabla C_{sin}|_{s}    
    \label{eq:A4}
\end{equation}
and the far-field boundary condition is $C_{d}|_{r\rightarrow \infty} = \, 0$.
\\

The solutions for $C_{sin}$ and $C_a$ [see Ref.~(1) for details] can be written in terms of Legendre polynomials $P_n[x]$ as
\begin{equation}
    C_{sin}(\bm{r}_p) 
    = C_0 \Tilde{\alpha} \frac{a}{|\bm{r}_p + \bm{R}|} = C_0 \Tilde{\alpha} \sum_{n\ge 0} (-1)^n \frac{a \,r_<^n}{r_>^{n+1}} P_n[\bm{\hat{R}} \cdot \bm{\hat{n}}]
    \label{eq:A5}
\end{equation}
and
\begin{equation}
    C_{a}(\bm{r}_p) = C_0 \sum_{n\ge 1 } \frac{K_n[\chi]}{(n+1)} \bigg(\frac{a}{r_p}\bigg)^{n+1} P_n[\bm{\hat{d}} \cdot \bm{\hat{n}}]
    \label{eq:A6} \, ,
\end{equation}
where $r_<$ and $r_>$ are chosen between $r_p$ and $R$,
$K_n[\chi] = (P_{n-1}[-\chi] - P_{n+1}[-\chi])/(1-\chi)$, $C_0 = a Q_e/ D$, and $\Tilde{\alpha} = \alpha/(4\pi a^2 Q_e)$.
We define $\bm{r}_p$ as the position vector with respect to the coordinate frame centered on the particle such that $\bm{r} = \bm{r}_p + \bm{R}$ and $\bm{r}_p|_{s} = a \bm{\hat{n}}$.
As $\bm{R}$ is constant for a given configuration, the gradient operator with respect to $\bm{r}_p$ and $\bm{r}$ are equal (i.e., $\nabla = \nabla_{\bm{r}} = \nabla_{\bm{r}_p}$).
To obtain a solution for $C_{d}$, we first rewrite the boundary condition in Eq.~(\ref{eq:A4}) in terms of Legendre polynomials as
\begin{equation}
\label{eq:A7}
\bm{\hat{n}} \cdot \nabla C_{d}|_{s} 
=  \frac{C_0\,\Tilde{\alpha}}{a} \sum_{n\ge 0} n  \bigg(\frac{-a}{R}\bigg)^{n+1} P_n[\bm{\hat{R}} \cdot \bm{\hat{n}}] \, .
\end{equation}
Equation~(\ref{eq:A7}) and the boundary condition at infinity suggests a solution for $C_{d}$ can be written in terms of $\bm{r}_p$ and Legendre polynomials of the form
\begin{equation}
C_{d}(\bm{r}_p) =  C_0 \Tilde{\alpha} \sum_{n \ge 0}  \frac{n (-1)^n}{n+1} \bigg(\frac{a^2}{R ~r_p}\bigg)^{n+1} P_n[\bm{\hat{R}} \cdot \bm{\hat{n}}].
\label{eq:A8}
\end{equation}
The solution for the total concentration is the sum of $C_a$, $C_{sin}$, and $C_d$ giving:
\begin{align}
\label{eq:A9}
C (\bm{r}) 
&= C_0 \sum_{n\ge 1 } \frac{K_n[\chi]}{(n+1)} \bigg(\frac{a}{r_p}\bigg)^{n+1} P_n[\bm{\hat{d}} \cdot \bm{\hat{n}}]
+ C_0 \Tilde{\alpha} 
\sum_{n\ge 0} (-1)^{n} \bigg(\frac{a~ r_<^n}{r_>^{n+1}}+\frac{n}{n+1} \bigg(\frac{a^2}{R~r_p}\bigg)^{n+1}\bigg) P_n[\bm{\hat{R}} \cdot \bm{\hat{n}}].
\end{align}

Prior to calculating the slip velocity via Eq.~(4) of the main text, we note that the phoretic mobility $b [\bm{\hat{n}}]$ for a spherical particle with bilateral symmetry can be expressed as
\begin{align}
b [\bm{\hat{n}}]/b_e = & 
\begin{cases}
1, ~~~~~~~~~~~~~~~ \bm{\hat{n}} \cdot \bm{\hat{d}} > -\chi\\
\beta, ~~~~~~~~~~~~~~~ \bm{\hat{n}} \cdot \bm{\hat{d}} < -\chi \\
\frac{1}{2} (1+\beta), ~~~~~ \bm{\hat{n}} \cdot \bm{\hat{d}} = -\chi
\end{cases} 
= 1 - (1-\beta) H[\bm{- \hat{n}} \cdot \bm{\hat{d}} - \chi]
\label{subeq:B3}
\end{align}
where $H(x)$ is the Heaviside function. 
The Heaviside function in Eq.~(\ref{subeq:B3}) can be expanded in terms of Legendre polynomials as
\begin{equation}
    \label{eq:B4}
    H[-\bm{\hat{n}} \cdot \bm{\hat{d}} - \chi]
    =\frac{1}{2} (1-\chi) - \frac{1}{2} (1-\chi) \sum_{n\ge 1} K_n[\chi] P_{n}[\bm{\hat{n}} \cdot \bm{\hat{d}}],
\end{equation}
Then, noting that $\bm{r}_p = \bm{r} - \bm{R}$ 
and $\nabla = \nabla_{\bm{r}} = \nabla_{\bm{r_p}}$, we have
\begin{equation}
\nabla (\bm{\hat{n}} \cdot \bm{\hat{d}})
= (\bm{\hat{d}} \cdot \nabla) \bm{\hat{n}} = (\bm{\hat{d}} \cdot \nabla_{\bm{r_p}}) \bm{\hat{n}} 
= \bm{\hat{d}} \cdot \big(\mathbb{I} - \bm{\hat{n}}\bm{\hat{n}}\big)
\label{eq:B5}
\end{equation}
and a similar relation can be derived for $\nabla (\bm{\hat{n}} \cdot \bm{\hat{R}})$.
Using these relations and the concentration field given by Eq.~(\ref{eq:A9}), the slip velocity can be expressed as
\begin{equation}
\bm{u}_{s}
 = - 2U_0\frac{b[\bm{\hat{n}}]}{b_e}\sum_{n\ge 1 } \frac{1}{n+1} \bigg(K_n[\chi]
P'_n[\bm{\hat{n}} \cdot \bm{\hat{d}}]\, \bm{\hat{d}}  
+ \Tilde{\alpha} (-1)^{n} (2n+1) \bigg(\frac{a}{R}\bigg)^{n+1} P'_n[\bm{\hat{n}} \cdot \bm{\hat{R}}]\, \bm{\hat{R}} \bigg)  \cdot \big(\mathbb{I} - \bm{\hat{n}}\bm{\hat{n}}\big)
\label{eq:B6}
\end{equation}
The total slip velocity is composed of two terms:
\begin{equation}
\bm{u}_{a} 
 = - 2U_0\frac{b[\bm{\hat{n}}]}{b_e}\sum_{n\ge 1 } \frac{1}{n+1} \bigg(K_n[\chi]
P'_n[\bm{\hat{n}} \cdot \bm{\hat{d}}]\, \bm{\hat{d}} \bigg)  \cdot \big(\mathbb{I} - \bm{\hat{n}}\bm{\hat{n}}\big)
\end{equation}
which is solely due to the intrinsic activity of the Janus particle and, 
\begin{equation}
\bm{u}_{sin}
 = - 2U_0\frac{b[\bm{\hat{n}}]}{b_e}\sum_{n\ge 1 } \frac{1}{n+1} \bigg(\Tilde{\alpha} (-1)^{n} (2n+1) \bigg(\frac{a}{R}\bigg)^{n+1} P'_n[\bm{\hat{n}} \cdot \bm{\hat{R}}]\, \bm{\hat{R}} \bigg)  \cdot \big(\mathbb{I} - \bm{\hat{n}}\bm{\hat{n}}\big)
\end{equation}
which arises due to the presence of the singularity.
\\

The translational velocity for a spherical Janus particle in an unbounded fluid is given by
\begin{equation}
\bm{U} = -\frac{1}{4 \pi} \int_{s} \bm{u}_{s} dS = -\frac{1}{4 \pi} \int_{s} [\bm{u}_{a} + \bm{u}_{sin}] dS  = \bm{U}_{a} + \bm{U}_{sin}.
\label{eq:B7}
\end{equation}
Assuming, without loss of generality, that $\bm{\hat{d}} = \bm{\hat{z}}$ such that  
$\bm{\hat{n}} = \sin\theta\cos\phi \bm{\hat{x}} + \sin\theta\cos\phi \bm{\hat{y}} + \cos\theta \bm{\hat{z}}$, 
the component of the translational velocity due to the intrinsic activity of the particle is given by
\begin{align}
\bm{U}_{a} = \frac{1}{2 \pi} U_0 \sum_{n\ge 1 } \frac{K_n[\chi]}{(n+1)}
\int_{0}^{2\pi} \int_{-1}^{1} \frac{b[\bm{\hat{n}}]}{b_e}  \big(\bm{\hat{d}} - \cos \theta \bm{\hat{n}} \big) 
P'_n[\cos \theta] d(\cos \theta) d\phi 
 = U_0 \bm{\hat{d}} \Big( 1+\chi - (1-\beta) B[\chi]\Big),
\label{eq:B8}
\end{align}
where 
\begin{equation}
B[\chi] =  \sum_{n\ge 1} \frac{K_n[\chi]}{(n+1)} \bigg( \big(1-\chi^2\big) P_n[-\chi] + \int_{-1}^{-\chi} 2 u P_n[u] du \bigg).
\label{eq:B9}
\end{equation}
Again without loss of generality, we assume that $\bm{\hat{R}} = \bm{\hat{z}}$, the component of the translational velocity due to the presence of the source can be written as
\begin{equation}
\bm{U}_{sin}  = \frac{1}{2 \pi} \Tilde{\alpha}U_0
\sum_{n\ge 1} (-1)^{n} \frac{2n+1}{n+1}   
\bigg(\frac{a}{R}\bigg)^{n+1}
\int_{0}^{2\pi} \int_{-1}^{1}\frac{b[\bm{\hat{n}}]}{b_e}~ \big(\bm{\hat{R}} - \bm{\hat{n}} \cos \theta \big)
P'_n[\cos \theta] d(\cos \theta) d\phi  
\label{eq:B10}
\end{equation}
Simplifying the integration in Eq.~(\ref{eq:B10}) to
\begin{align}
\label{eq:B11}
& \int_{0}^{2\pi}\int_{-1}^{1} \frac{b[\bm{\hat{n}}]}{b_e} ~\big(\bm{\hat{R}} - \bm{\hat{n}} \cos \theta \big)  P'_n[\cos \theta] d(\cos \theta) d\phi \nonumber \\
& = \int_{0}^{2\pi}\int_{-1}^{1} \big(\bm{\hat{R}} - \bm{\hat{n}} \cos \theta \big)  P'_n[\cos \theta] d(\cos \theta) d\phi \nonumber \\
& - (1-\beta) \int_{0}^{2\pi}\int_{-1}^{1} H[-\bm{\hat{n}} \cdot \bm{\hat{d}} - \chi] \big(\bm{\hat{R}} - \bm{\hat{n}} \cos \theta \big)  P'_n[\cos \theta] d(\cos \theta) d\phi \nonumber \\
& = \Big(1 - \frac{1}{2} (1-\beta) (1-\chi) \Big) \frac{8 \pi}{3} \delta_{n,1} \bm{\hat{R}}  \nonumber \\
& + \frac{1}{2} (1-\beta) (1-\chi) \sum_{l\ge 1} K_l[\chi]  \int_{0}^{2\pi}\int_{-1}^{1} P_{l}[\bm{\hat{n}} \cdot \bm{\hat{d}}] \big(\bm{\hat{R}} - \bm{\hat{n}} \cos \theta \big)  P'_n[\cos \theta] d(\cos \theta) d\phi.
\end{align}
The last line follows from applying the Heaviside function expansion given in Eq.~(\ref{eq:B4}).
We are able to further simplify Eq.~(\ref{eq:B11}) and find its projection along $\bm{\hat{R}}$ and $\bm{\hat{\varphi}}$ by expanding $P_l[\bm{\hat{n}} \cdot \bm{\hat{d}}]$ in terms of associated Legendre polynomials $P_{n}^{m}[x]$.
If we consider the general case where $\bm{R}$ and $\bm{\hat{d}}$ are not in the same plane such that $\bm{\hat{d}}$ and $\bm{\hat{n}}$ have polar and azimuthal angles $\Theta$ and $\phi'$, and $\theta$ and $\phi$, respectively, i.e., $\bm{\hat{d}} = \cos\Theta \bm{\hat{z}} + \sin\Theta \cos\phi' \bm{\hat{x}} + \sin\Theta \sin\phi' \bm{\hat{y}}$, and $\bm{\hat{n}} = \cos\theta \bm{\hat{z}} + \sin\theta\cos\phi \bm{\hat{x}} + \sin\theta\sin\phi \bm{\hat{y}}$, then 
\begin{align}
\label{eq:B12}
P_l[\bm{\hat{n}} \cdot \bm{\hat{d}}]  
= \sum_{m=-l}^{l} \frac{(l-m)!}{(l+m)!} P_{l}^{m}[\cos\Theta] P_{l}^{m}[\cos\theta]e^{-im\phi'} e^{im\phi} \, .
\end{align}
Upon substitution of Eq.~(\ref{eq:B12}) into Eq.~(\ref{eq:B11}), we obtain
\begin{align}
\label{eq:B13}
&\int_{0}^{2\pi}\int_{-1}^{1} P_{l}[\bm{\hat{n}} \cdot \bm{\hat{d}}] \big(\bm{\hat{R}} - \bm{\hat{n}} \cos \theta \big)  P'_n[\cos \theta] d(\cos \theta) d\phi \nonumber \\
& =  \sum_{m=-l}^{l} \frac{(l-m)!}{(l+m)!} P_l^m[\cos\Theta] e^{-i m\phi'} 
\int_{0}^{2\pi}\int_{-1}^{1} P_l^m[\cos\theta] \big(\bm{\hat{R}} - \bm{\hat{n}} \cos \theta \big) P'_n[\cos \theta] e^{i m \phi} d(\cos \theta) d\phi 
\nonumber \\
& =  2\pi P_l[\cos\Theta] \int_{-1}^{1} \big(1-\cos^2\theta\big)P_l[\cos \theta]P'_n[\cos \theta]d(\cos \theta) \bm{\hat{R}} \nonumber \\
&\quad   -\frac{(l-m)!}{(l+m)!} P_l^m[\cos\Theta] e^{-i m\phi'} \pi 
\int_{-1}^{1} \cos\theta \sin\theta P_l^m[\cos\theta] P'_n[\cos \theta] d(\cos\theta) (\delta_{m,1} + \delta_{m,-1})  \big(\bm{\hat{x}}+im\bm{\hat{y}}\big)\nonumber \\
& =  2\pi P_l[\cos\Theta] \frac{n(n+1)}{2n+1} \frac{2}{2l+1}(\delta_{l,n-1}-\delta_{l,n+1}) \bm{\hat{R}}
+ 2\pi P_l^{1}[\cos\Theta]
\frac{1}{2n+1}\frac{2}{2l+1} \big(n \delta_{l,n+1} + (n+1)\delta_{l,n-1} \big) \bm{\hat{\varphi}}.
\end{align}
We can now write $\bm{U}_{sin}$ in terms of $\bm{R}$ and $\bm{\varphi}$ by substituting Eq.~(\ref{eq:B13}) into Eq.~(\ref{eq:B11}) and substituting the resulting expression into Eq.~(\ref{eq:B10}) to obtain
\begin{align}
\label{eq:B14}
\bm{U}_{sin}
& = U_0 \Tilde{\alpha} 
\Big(- 2\frac{a^2}{R^2} + (1-\beta) M[\chi,\bm{R},\bm{\hat{d}}] \Big)  \bm{\hat{R}} 
- U_0 \Tilde{\alpha} (1-\beta) \sin\Theta N[\chi,\bm{R},\Theta] \bm{\hat{\varphi}},
\end{align}
where, $M[\chi,\bm{R},\Theta]$ and $N[\chi,\bm{R},\Theta]$ are given by
\begin{subequations}
\label{eq:B15}
\begin{align}
\label{subeq:B15a}
M[\chi,\bm{R},\Theta] = 
 \frac{a^2}{R^2} (1-\chi) \bigg(1- \frac{1}{4}\chi(1+\chi)(3\cos^2\Theta-1) \bigg)
 & + (1-\chi)\sum_{n\ge 2} \bigg(\frac{a}{R}\bigg)^{n+1} 
\Big(n A_{n+1} - n A_{n-1}\Big),
\end{align}
\begin{align}
\label{subeq:B15b}
N[\chi,\bm{R},\Theta] &=
-\frac{3}{4}  \frac{a^2}{R^2}\chi (1-\chi^2) \cos\Theta
- (1-\chi)\sum_{n\ge 2} 
\frac{1}{n+1} \bigg(\frac{a}{R}\bigg)^{n+1} \frac{d}{d(\cos\Theta)} \bigg(n A_{n+1}+ (n+1) A_{n-1}\bigg)\, ,
\end{align}
and
\[A_{n} = \frac{1}{2n+1} (-1)^n P_{n}[\cos\Theta]K_{n}[\chi]. \]
\end{subequations}
The total translation velocity [Eq.~(7a,b) of the main text] is given by $\bm{U} = \bm{U}_{a} + \bm{U}_{sin}$.
\\

Prior to calculating the angular velocity of a particle in the presence of a point singularity, we recapitulate a useful result derived in Ref.~\cite{popescuchemotaxis18}. 
With the aid of the vector calculus identity $\int_{s} \bm{\hat{n}} \times \nabla (b[\bm{\hat{n}}] C) ~dS = \int \nabla \times \nabla (b[\bm{\hat{n}}] C) dV =0$, we can write the angular velocity as
\begin{align}
\label{eq:B16}
\bm{\Omega} & = -\frac{3}{8 \pi a} \int_{s} \bm{\hat{n}} \times \bm{u}_{s} dS \nonumber \\
&= - \frac{3}{8 \pi a} \int_{s} -\bm{\hat{n}} \times b[\bm{\hat{n}}] \nabla C  ~dS \nonumber \\
&= - \frac{3}{8 \pi a} \int_{s} C ~ \bm{\hat{n}} \times \nabla b[\bm{\hat{n}}] + \frac{3}{8 \pi a} \int_{s} \bm{\hat{n}} \times \nabla (b[\bm{\hat{n}}] C)  ~dS \nonumber \\
& = - \frac{3}{8 \pi a^2} b_e (1-\beta) \int_{s} ( C \delta[\bm{\hat{n}} \cdot \bm{\hat{d}} + \chi]~ \bm{\hat{n}} \times \bm{\hat{d}} ) ~dS \nonumber \\
&= \frac{3}{8 \pi a^2} b_e (1-\beta) \int_{\bm{\hat{n}} \cdot \bm{\hat{d}} = -\chi} C ~\bm{\hat{d}} \times \bm{\hat{n}} ~dl.
\end{align}
Interestingly, the angular velocity of a spherical particle with bilateral Janus symmetry is proportional to $1-\beta$ and is simply the vector cross product $C ~\bm{\hat{d}} \times \bm{\hat{n}}$ along the contour separating the emitting and absorbing region of the particle (i.e., $\bm{\hat{n}} \cdot \bm{\hat{d}} = -\chi$). 
It is straightforward to show for spherical particle with bilateral symmetry the contribution to the angular velocity from the intrinsic activity of the particle is zero:
\begin{align}
\label{eq:B18}
\bm{\Omega}_{a} & =- \frac{3}{8 \pi a^2} b_e(1-\beta) \int_{s} C_{a} ~\delta[\bm{\hat{n}} \cdot \bm{\hat{d}} + \chi]~ \bm{\hat{n}} \times \bm{\hat{d}} ~d\Omega \nonumber \\
& = - \frac{3}{4 \pi a} U_0 (1-\beta) \sum_{n\ge 1 } \frac{(1-\chi) K_n[\chi]}{n+1} P_n(-\chi)  \int_{s} \delta[\bm{\hat{n}} \cdot \bm{\hat{d}} + \chi]~ \bm{\hat{n}} ~d\Omega \times \bm{\hat{d}}= 0.
\end{align}
This results of $\bm{\Omega}_{a} = 0$ is valid for all phoretic mobility ratios $\beta$ and Janus balances $\chi$. 
The contribution to the angular velocity due to the presence of the singularity is given by
\begin{align}
\label{eq:B19}
\bm{\Omega}_{sin} & = - \frac{3}{8 \pi a^2} b_e(1-\beta) \int_{s} (C_{sin} + C_{d}) ~\delta[\bm{\hat{n}} \cdot \bm{\hat{d}} + \chi]~ \bm{\hat{n}} \times \bm{\hat{d}} ~d\Omega \nonumber \\
& = - \frac{3}{4 \pi a} U_0(1-\beta) \Tilde{\alpha} 
\sum_{n\ge 0} (-1)^{n}\frac{2n+1}{n+1} \bigg(\frac{a}{R}\bigg)^{n+1}
\int_{s} P_n[\bm{\hat{R}} \cdot \bm{\hat{n}}]\delta[\bm{\hat{n}} \cdot \bm{\hat{d}} + \chi]~ \bm{\hat{n}} ~d\Omega \times \bm{\hat{d}}.
\end{align}
The integration in Eq.~(\ref{eq:B19}) can be simplified using a similar procedure as done for the translation velocity, where we first assume, without loss of generality, that $\bm{\hat{d}} = \bm{\hat{z}}$, and thus $\bm{\hat{n}} = \cos\theta \bm{\hat{z}}+ \sin \theta \cos \phi \bm{\hat{x}} + \sin \theta \sin \phi \bm{\hat{y}}$ 
and $\bm{\hat{R}} = \cos \Theta ~\bm{\hat{z}} + \sin \Theta \cos \phi' ~\bm{\hat{x}} + \sin \Theta \sin \phi' ~\bm{\hat{y}}$. 
We then write $P_n[\bm{\hat{R}} \cdot \bm{\hat{n}}]$ in terms of associated Legendre polynomials $P_{n}^{m}[x]$ where the integration in Eq.~(\ref{eq:B19}) simplifies to
\begin{align}
\label{eq:B21}
&\int_{s} P_n[\bm{\hat{R}} \cdot \bm{\hat{n}}]\delta[\bm{\hat{n}} \cdot \bm{\hat{d}} +\chi]~ \bm{\hat{n}} ~d\Omega \times \bm{\hat{d}}  \nonumber \\
& = \sum_{m=-n}^{n}  
\frac{(n-m)!}{(n+m)!} P_n^m[\cos \Theta]  e^{-i m \phi'} \int_0^{2\pi} \int_0^\pi e^{i m \phi} P_n^m[\cos\theta] \bm{\hat{n}} \delta[\cos\theta+\chi] \sin\theta d\theta d\phi
 \times \bm{\hat{d}} \nonumber \\
& = \sum_{m=-n,n\neq0}^{n} \pi \, (1-\chi^2)
 \frac{(n-m)!}{(n+m)!} P_n^m[\cos \Theta] P_n^m[-\chi] e^{-i m \phi'}
\bigg((\delta_{m,1} + \delta_{m,-1}) (\bm{\hat{x}} + i m \bm{\hat{y}}) \bigg)  \times \bm{\hat{d}} \nonumber \\
& =  2\pi (1-\chi^2)\frac{(n-1)!}{(n+1)!} P_n^1[\cos \Theta] P_n^1[-\chi] \big(\cos \phi' \bm{\hat{x}} + \sin \phi'  \bm{\hat{y}} \big)  \times \bm{\hat{d}} \nonumber \\
& = \frac{2\pi (1-\chi) \sqrt{1-\chi^2} }{2n+1} \sin\Theta \frac{dP_{n}[\cos \Theta]}{d(\cos\Theta)} K_n[\chi] ~\frac{\bm{\hat{R}} \times \bm{\hat{d}}}{|\bm{\hat{R}} \times \bm{\hat{d}}|}.
\end{align}
The integration result from Eq.~(\ref{eq:B21}) can be substituted into Eq.~(\ref{eq:B19})  to obtain the expression in the main text [Eq.~(7c)] for the total angular velocity:
\begin{align}
\label{eq:B22}
\bm{\Omega} 
& = - \frac{3 U_0\Tilde{\alpha}}{2 a}(1-\beta)\sin\Theta ~ \omega[\chi,\bm{R},\bm{\hat{d}}] ~\frac{\bm{\hat{R}} \times \bm{\hat{d}}}{|\bm{\hat{R}} \times \bm{\hat{d}}|},
\end{align}
where
\begin{equation}
\label{eq:B23}
\omega[\chi,\bm{R},\Theta] = \sqrt{1-\chi^2} (1-\chi)\sum_{n\ge 1} (-1)^{n}\frac{1}{n+1} \bigg(\frac{a}{R}\bigg)^{n+1} \frac{dP_{n}[\cos \Theta]}{d(\cos\Theta)} K_n[\chi].
\end{equation}
Lastly, we note the angular velocity is only non-zero in the direction perpendicular to $\bm{\hat{d}}$ and $\bm{R}$, restricting the motion of the particle to the two dimensional plane containing the particle orientation $\bm{\hat{d}}$ and radial vector $\bm{\hat{R}}$.  

\clearpage

\subsubsection{Radial dependence of dimensionless parameters $B[\chi]$, $M[\chi, R, \Theta], N[\chi, R, \Theta]$, and $\omega[\chi, R, \Theta]$ for $\chi = 0$}
\phantom{a}

\begin{figure*}[h!]
	\centering
        \includegraphics[width=1.\textwidth]{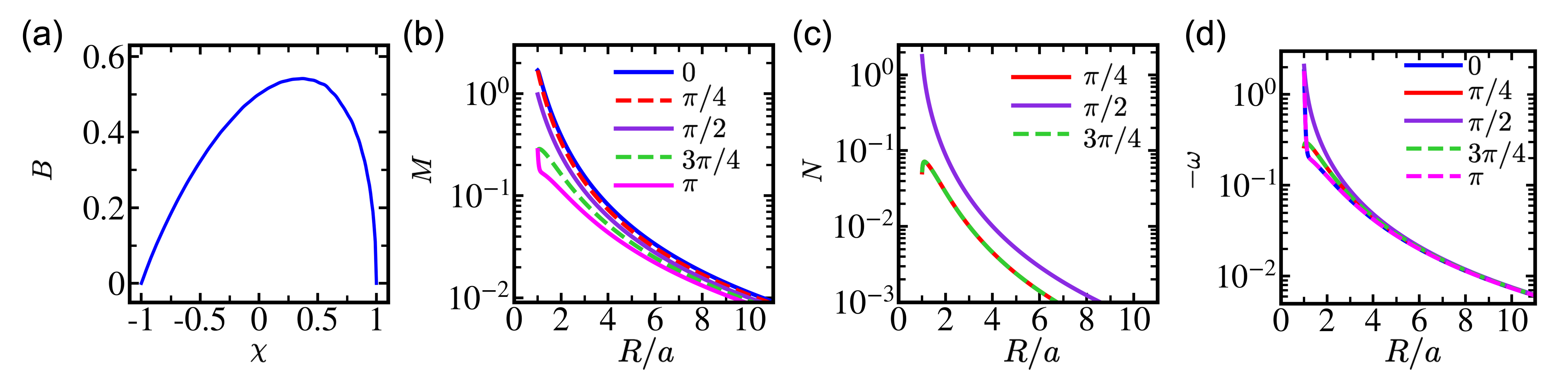}
	\caption{\protect\small{
(a) Parameter $B[\chi]$ as a function of Janus-balance $\chi$. (b--d) Parameters $M[\chi, R, \Theta]$, $N[\chi, R, \Theta]$, and $\omega[\chi, R, \Theta]$ for $\chi = 0$ for different orientations $\Theta$ as a function of the radial distance from the singularity. Note that $N[\chi, R, \Theta]$ vanishes for $\Theta=0$ and $\pi$. 
 }}
	\label{fig:figure_S1}
\end{figure*}
  
\subsubsection{Location of fixed points as a function of the strength of the singularity}
\begin{figure*}[ht!]
	\centering
        \includegraphics[width=0.75\textwidth]{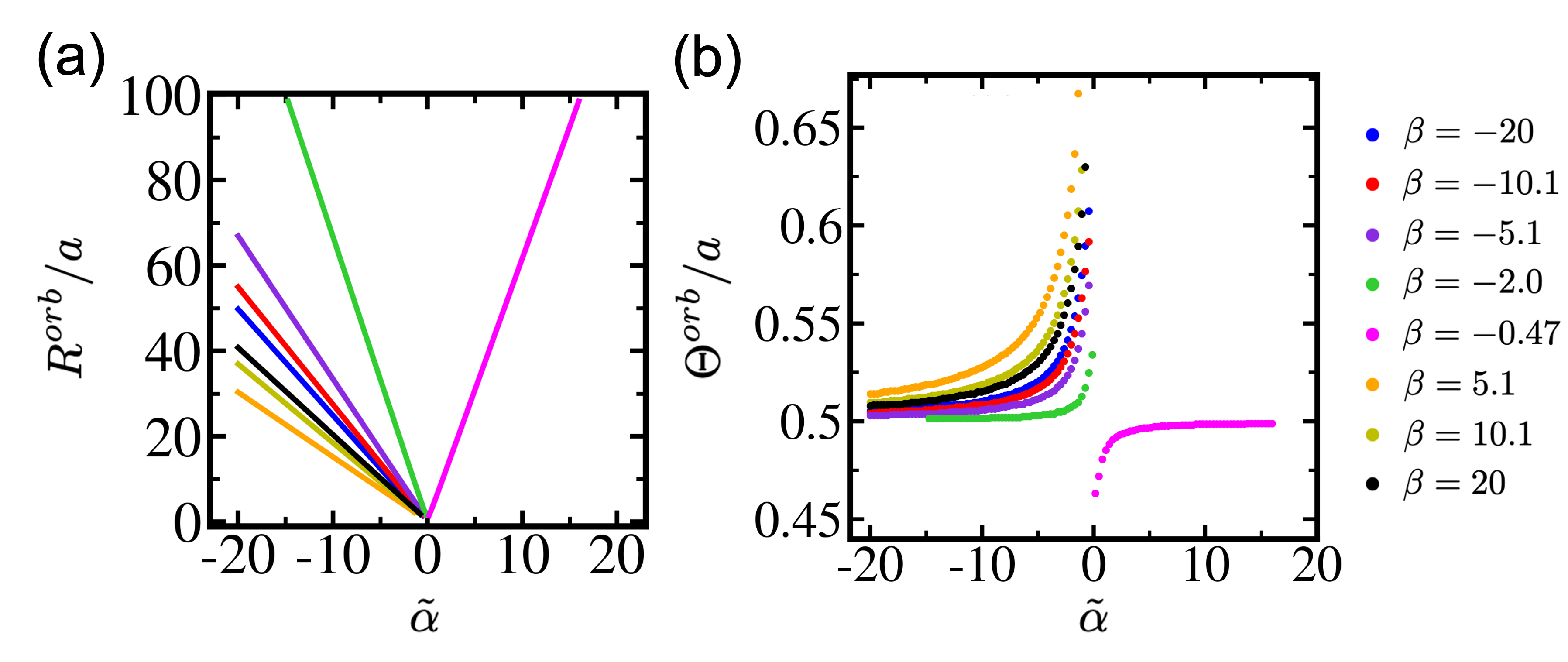}
	\caption{\protect\small{
 The radial $R^{orb}/a$ and orientation $\Theta^{orb}$ dependence of orbiting states as a function of the strength of the singularity $\Tilde{\alpha}$ for different phoretic mobility ratios $\beta$. As the strength of the singularity increases, orbiting states occur further from the singularity, and the orientation asymptotically approaches an angle of $\Theta^{orb} = \pi/2$. }}
	\label{fig:figure_S3}
\end{figure*}
\begin{figure*}[ht!]
	\centering
        \includegraphics[width=1.\textwidth]{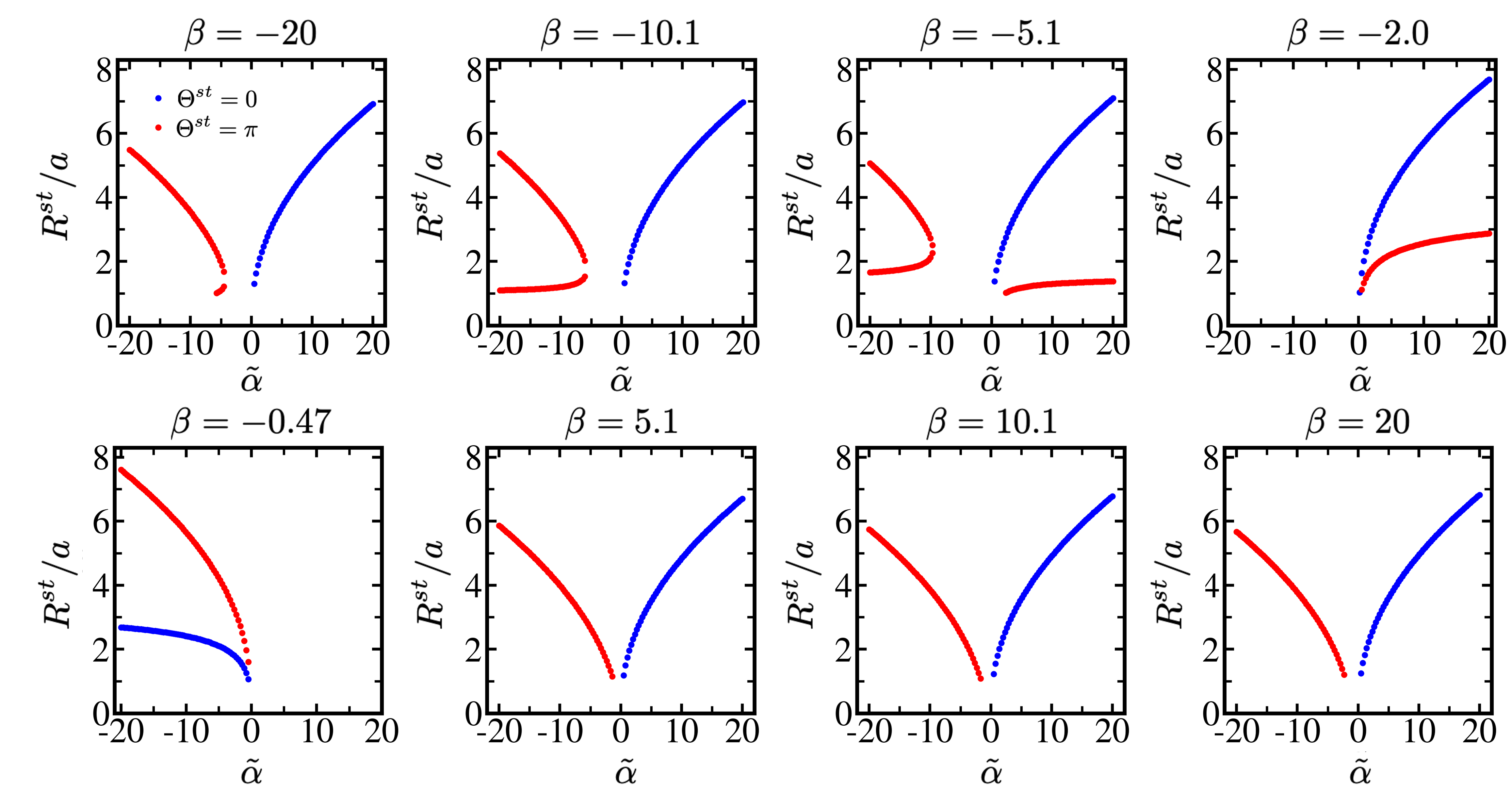}
	\caption{\protect\small{The radial dependence of stationary states $R^{st}/a$ with orientations $\Theta^{st} = 0$ (blue) and $\pi$ (red) as a function of the strength of the singularity $\Tilde{\alpha}$ for various phoretic mobility ratios $\beta$. 
The general trend is that as the singularity's strength increases, the fixed points occur further away from the singularity.
For $\Theta=0$, stationary states occur exclusively in the presence of a point source ($\Tilde{\alpha} > 0$) for all phoretic mobility ratios $\beta$, except in interval $-1<\beta<0$ where the stationary state arise in the presence of a point source.
For $\beta=-10.1$ and $\beta=-5.1$, there are two stationary states for $\Theta=\pi$ for a broad range of $\Tilde{\alpha}$. As the strength of the sink decreases, these two branches of stationary states converge to a single stationary stare before disappearing entirely.}}
	\label{fig:figure_S2}
\end{figure*}

\hphantom{invisible}
\clearpage
\subsubsection{Details of Simulation Videos}
Each movie shows the real-space trajectory for the different values of the phoretic mobility ratio in Figures 3, 4, and 5 of the main text. The trajectory colors are consistent with those shown in the figures, and the vector corresponds to the orientation $\Theta$. The singularity is represented as a grey circle, and the time is given in units of $a/U_0 = 2 a D/(Q_e b_e)$.
\begin{enumerate}
    \item \texttt{video\_1(fig4a).mp4} -- $U_0=1$, $\Tilde{\alpha}=10$ and $\beta=-2$
    \vspace{-0.4cm}
    \item \texttt{video\_2(fig4b).mp4} -- $U_0=1$, $\Tilde{\alpha}=10$ and $\beta=0$
    \vspace{-0.4cm}
    \item \texttt{video\_3(fig4c).mp4} -- $U_0=1$, $\Tilde{\alpha}=10$ and $\beta=5$
    \vspace{-0.4cm}
    \item \texttt{video\_4(fig5a).mp4} -- $U_0=1$, $\Tilde{\alpha}=-10$ and $\beta=-15$
    \vspace{-0.4cm}
    \item \texttt{video\_5(fig5b).mp4} -- $U_0=1$, $\Tilde{\alpha}=-10$ and $\beta=-2$
    \vspace{-0.4cm}
    \item \texttt{video\_6(fig5c).mp4} -- $U_0=1$, $\Tilde{\alpha}=-10$ and $\beta=10$
    \vspace{-0.4cm}
    \item \texttt{video\_7(fig6).mp4} -- $U_0=-1$, $\Tilde{\alpha}=10$ and $\beta=0.644$
\end{enumerate}


\providecommand{\noopsort}[1]{}\providecommand{\singleletter}[1]{#1}
\providecommand{\latin}[1]{#1}
\makeatletter
\providecommand{\doi}
  {\begingroup\let\do\@makeother\dospecials
  \catcode`\{=1 \catcode`\}=2 \doi@aux}
\providecommand{\doi@aux}[1]{\endgroup\texttt{#1}}
\makeatother
\providecommand*\mcitethebibliography{\thebibliography}
\csname @ifundefined\endcsname{endmcitethebibliography}  {\let\endmcitethebibliography\endthebibliography}{}